%% file: paper.tex
\documentclass[sigconf]{acmart}

\usepackage{color}

\usepackage{multirow}
\usepackage{colortbl}
\newcommand\Tstrut{\rule{0pt}{2.2ex}} % = `top' strut

\usepackage{balance}

\newcommand{\citeauth}[1]{\citeauthor{#1}~\cite{#1}} % ben

\usepackage{tikz}
\newcommand\copyrightnotice[1]{
	\begin{tikzpicture}[remember picture,overlay]
		\node[anchor=south,yshift=20pt] at (current page.south) {\fbox{\parbox{\dimexpr\textwidth-\fboxsep-\fboxrule\relax}{#1}}};
	\end{tikzpicture}
}

\AtBeginDocument{%
  \providecommand\BibTeX{{%
    \normalfont B\kern-0.5em{\scshape i\kern-0.25em b}\kern-0.8em\TeX}}}

\copyrightyear{2021}
\acmYear{2021}
\setcopyright{acmlicensed}\acmConference[CYSARM '21]{Proceedings of the 3rd Workshop on Cyber-Security Arms Race}{November 19, 2021}{Virtual Event, Republic of Korea}
\acmBooktitle{Proceedings of the 3rd Workshop on Cyber-Security Arms Race (CYSARM '21), November 19, 2021, Virtual Event, Republic of Korea}
\acmPrice{15.00}
\acmDOI{10.1145/3474374.3486915}
\acmISBN{978-1-4503-8661-6/21/11}

\begin{document}
\fancyhead{}

\title[A Study on Collaborative Machine Learning for DGA Detection]{The More, the Better? \\ A Study on Collaborative Machine Learning for DGA Detection}

\author{Arthur Drichel}
\email{drichel@itsec.rwth-aachen.de}
\affiliation{%
	\institution{RWTH Aachen University}
	\city{}
	\country{}
	%	\city{Aachen}
	%	\country{Germany}
}
\author{Benedikt Holmes}
\email{holmes@itsec.rwth-aachen.de}
\affiliation{
	\institution{RWTH Aachen University}
	\city{}
	\country{}
}
\author{Justus von Brandt}
\email{justus.von.brandt@rwth-aachen.de}
\affiliation{
	\institution{RWTH Aachen University}
	\city{}
	\country{}
}
\author{Ulrike Meyer}
\email{meyer@itsec.rwth-aachen.de}
\affiliation{%
	\institution{RWTH Aachen University}
	\city{}
	\country{}
}

\renewcommand{\shortauthors}{Drichel et al.}

\input{content/abstract.tex}






\maketitle

\copyrightnotice{\copyright\space Copyright held by the owner/author(s) 2021. This is the author's version of the work. It is posted here for your personal use. Not for redistribution. The definitive version was published in Proceedings of the 3rd Workshop on Cyber-Security Arms Race
(CYSARM ’21), https://doi.org/10.1145/3474374.3486915}

\input{content/introduction.tex}
\input{content/related_work.tex}
\input{content/evaluation_setup.tex}
\input{content/evaluation.tex}
\input{content/privacy_discussion.tex}
\input{content/conclusion.tex}

\begin{acks}
	The authors would like to thank Daniel Plohmann for granting
	us access to DGArchive as well as Jens Hektor from the IT Center of RWTH Aachen University, Masaryk University, CESNET, and Siemens AG for providing NXD data. 
	This project has received funding from the European Union's Horizon 2020 research and innovation programme under grant agreement No 833418. 
	Simulations were performed with computing resources granted by RWTH Aachen University under project rwth0438.
\end{acks}

%\vfill\eject

\bibliographystyle{ACM-Reference-Format}
\balance
\bibliography{bibliography}

%\appendix
%\input{content/appendix.tex}

\end{document}

%% file: content/abstract.tex
% !TEX root = ../paper.tex
\begin{abstract}
Domain generation algorithms (DGAs) prevent the connection between a botnet and its master from being blocked by generating a large number of domain names. 
Promising single-data-source approaches have been proposed for separating benign from DGA-generated domains.
Collaborative machine learning (ML) can be used in order to enhance a classifier's detection rate, reduce its false positive rate (FPR), and to improve the classifier's generalization capability to different networks.
In this paper, we complement the research area of DGA detection by conducting a comprehensive collaborative learning study, including a total of 13,440 evaluation runs.
In two real-world scenarios we evaluate a total of eleven different variations of collaborative learning using three different state-of-the-art classifiers.
We show that collaborative ML can lead to a reduction in FPR by up to 51.7\%. However, while collaborative ML is beneficial for DGA detection, not all approaches and classifier types profit equally.
We round up our comprehensive study with a thorough discussion of the privacy threats implicated by the different collaborative ML approaches.
\end{abstract}

%% file: content/introduction.tex
% !TEX root = ../paper.tex
\section{Introduction}
\label{sec:introduction}

Domain generation algorithms (DGAs) are used by botnets to establish a connection between infected hosts and their command and control (C2) server.
To this end, DGAs generate and query a large number of domain names, most of which result in non-existing domain (NXD) responses.
This is because the botnet master only registers a handful of the generated domains while the bots query each domain.
As a result, DGAs create an asymmetrical situation in which defenders have to block any generated domain name while a single resolving domain is sufficient for the botnet to receive new commands.

As a countermeasure, several machine learning (ML) classifiers have been proposed that attempt to separate benign from DGA-generated domains.
These classifiers are trained in a supervised manner using malicious and benign labeled samples.
Malicious labeled data can be easily obtained from open source intelligence (OSINT) feeds such as DGArchive~\cite{plohmann_comprehensive_2016}.
Domain names that are available in public lists such as Alexa\footnote{https://www.alexa.com/topsites} or Tranco~\cite{lepochat_tranco_2019} can be used as benign data.
However, it has been shown that classifiers trained on public benign data are less robust against adversarial attacks~\cite{drichel_analyzing_2020}.
Moreover, in Section~\ref{sec:sharing_approaches}, we show that training with publicly available data is insufficient for DGA detection on private NXD data.
There are additional practical benefits of using NXDs extracted from non-resolving DNS traffic (NX-traffic) for classifier training.
First, DGAs are typically recognized in NX-traffic long before they resolve a registered domain for their C2 server.
Therefore, infected hosts can be disinfected before they are ordered to take part in malicious actions. 
Second, the amount of NX-traffic is an order of magnitude less than the amount of full DNS traffic, which makes it easier to monitor.
And third, NXDs are less privacy sensitive compared to resolving domain names because it is not possible to recover the full browser history of users from NXDs.

Nevertheless, NXDs may also contain sensitive information about individuals within an organization or confidential information about an organization as a whole.
For instance, typing errors can still hint websites visited by employees.
Moreover, the knowledge of NXDs generated by misconfigured or outdated software could be leveraged by an adversary to attack the organization's network.
Due to these privacy issues, it is obviously not possible to directly share benign NXDs in a collaborative ML setting for DGA detection.
Thus, we consider the disclosure of benign samples to be the main privacy-critical aspect in our use case.

The goal of collaborative ML is to solve a data-driven task by using data stored by several different parties, on the assumption that more can be learned by combining knowledge from the participants' local datasets.
Thereby, collaborative learning can aid in generalization of a trained global model and is a vital tool whenever local datasets have significant pairwise statistical difference or comprise only few samples each.
In the past, several approaches for collaborative learning have been proposed including: (1) Transfer Learning,
(2) Ensemble constructs, (3) Teacher-Student (T/S) learning, and (4) Federated Learning (FL). 
In each approach, different types of intelligence are shared including trained models or parts thereof, predictions for unknown samples, or weight updates during training.
Depending on the type of intelligence shared, the precision with which information on the sensitive training data may be inferred varies.

In the context of DGA detection many single-data-source approaches exist.
This paper complements this research area with an analysis of collaborative learning on this task.
To the best of our knowledge, such a study has not been done so far and is missing in related literature.

Our contribution primarily focuses on a comparative evaluation of different collaborative learning approaches using three different state-of-the-art classifiers.
Thereby, we answer the following three research questions: 
\begin{enumerate}
	\item Is collaborative training beneficial for organizations that mostly classify data from their own network? 
	\item Do jointly trained classifiers improve in their detection performance for samples originating from different networks?
	\item Do jointly trained classifiers improve in their detection performance with increasing participants?
\end{enumerate}
The first two research questions are derived from possible real-world application environments for trained classifiers.
The answer to the last research question reveals suitable approaches for DGA detection.
In fact, we show that out of the four investigated collaborative ML approaches only Feature Extractor Sharing (an approach related to Transfer Learning) and FL are beneficial for DGA detection.
Four collaborative ML variants based on T/S learning and Ensemble classification lead even to worse results compared to the baseline.
In addition, in our study we compare three different state-of-the-art classifiers and show that different types of classifiers are better suited for different sharing approaches. 

Secondary, we thoroughly discuss the privacy aspects concerning the disclosure of benign training samples for the collaborative approaches Feature Extractor Sharing and FL as well as the performance implications caused by privacy-preserving measures.
As data is never directly shared in our approaches, we link data privacy solely to the trained models' or training procedures' vulnerability to privacy-threatening inference attacks.

%% file: content/related_work.tex
% !TEX root = ../paper.tex
\section{Related Work}
\label{sec:related_work}
In the following we briefly present the studied use case, introduce collaborative ML approaches as well as provide an overview on privacy research in deep learning (DL).

\subsection{DGA Detection}
Various approaches have been proposed in the past to capture DGA activity within networks.
These approaches can be roughly divided into context-less (e.g.,~\cite{schuppen_fanci_2018,woodbridge_predicting_2016,yu_character_2018,saxe_expose_2017,drichel_analyzing_2020}) and context-aware approaches (e.g.,~\cite{antonakakis_throwaway_2012,bilge_exposure_2014, grill_detecting_2015,yadav_winning_2012, schiavoni_phoenix_2014,shi_malicious_2018}). 
The former group uses information extracted only from individual domain names and ignores any contextual data to separate benign from DGA-generated domains.
On the other hand, context-aware approaches use additional knowledge to further improve the detection performance.
Previous studies such as \cite{drichel_analyzing_2020,woodbridge_predicting_2016,schuppen_fanci_2018,yu_character_2018} have shown that the context-less approaches achieve state-of-the-art performance, are less resource-demanding, and are less privacy-invasive than context-aware approaches.

The proposed context-less ML classifiers can be further divided into feature-based such as support vector machines or random forests (e.g.,~\cite{schuppen_fanci_2018}), and feature-less (DL) classifiers such as recurrent (RNN), convolutional (CNN), or residual neural networks (ResNet) (e.g.,~\cite{woodbridge_predicting_2016,yu_character_2018,saxe_expose_2017,drichel_analyzing_2020}).
When comparing both types of classifiers, it was shown that the approaches based on DL achieve superior detection performance~\cite{drichel_analyzing_2020,woodbridge_predicting_2016,peck_charbot_2019,spooren_detection_2019}.

All of these proposed approaches have in common that they are single-data-source approaches.
In this work, we use context-less DL based classifiers trained on NX-traffic to complement this research area with an analysis of collaborative learning for this detection task.

\subsection{Collaborative Machine Learning}
Focusing on different attributes such as communication \& computation costs, retraining effort, as well as data availability, locality or privacy, yields various sharing approaches. We list the following approaches by an increasing order regarding the involvement and intercommunication of the sharing parties.
Related collaborative learning or sharing approaches include (1) Transfer Learning, (2) Ensemble constructs, as well as more involved approaches, such as (3) Teacher-Student learning and (4) Federated Learning.

In Transfer Learning (e.g.,~\cite{pan_survey_2010, goodfellow_deep-learning_2016, weiss_survey_2016, zhuang_comprehensive_2021} and literature bodies referenced therein) a consuming party may leverage an existing neural network model pre-trained by a different party on a related task using a larger, more general, or more complex dataset than the one owned by the consumer. The consuming party can exchange and fit the tail of the model (decision layers) to its own dataset, making use of the knowledge held in the weights of the pre-trained extraction layers which remain unchanged during retraining.

Ensemble classifiers (e.g.,~\cite{dietterich_ensemble_2000, goodfellow_deep-learning_2016, sagi_ensemble_2018}) denote a collection of ML models (trained on the same task) with the advantage that the individual errors of each single model are rectifiable by the others. Ensembles can be trained as a collective or constructed from pre-trained models to provide a global inference interface by combination of their distinct outputs (e.g., by averaging soft-labels or voting). In this work we view the latter case.

In a Teacher-Student (T/S) scenario (originated from \cite{hinton_distilling_2015}) one or multiple pre-trained teacher models guide the training of a student model.
Guided training can, for instance, serve the purpose of improved training time on related tasks or model compression while maintaining the teacher's behavior and utility on the certain task (a concept known as knowledge distillation \cite{hinton_distilling_2015}). In our work, we investigate the student's capability to retain and combine intelligence from all teachers.

Federated Learning (FL), introduced by \citeauth{mcmahan_communication-efficient_2017}, enhances the state of distributed collaborative learning, as participants are not required to directly share their local private data but instead share gradient updates in an iterative training process in which local updates are aggregated and applied to a global model.

Impact of collaborative learning and sharing approaches with regard to improved performance in DL can be witnessed by examples in the research fields computer vision and natural language processing (e.g., \cite{razavian_cnn-offtheshelf_2014, hinton_distilling_2015, yang_applied_2018}).

\subsection{Privacy in Machine Learning}
Due to its high consummation of sensitive data, DL models or training procedures have become the main suspect of research investigating threats (termed inference attacks) and defenses concerning the natural information leakage of consumed training data that is inherent to learning (e.g., \cite{ateniese_hacking_2015, fredrikson_model-inversion_2015, abadi_dpsgd_2016, shokri_membership-inference_2017, papernot_sok_2018, al_rubaie_ppml_2019, nasr_comprehensive_2019}).

The most prominently researched privacy threat against classification models is the Membership Inference attack (MemI) \cite{shokri_membership-inference_2017} which aims at disclosing the participation of a known data sample in the training process of the targeted model. Another relevant attack is Model Inversion \cite{fredrikson_warfarin_2014, fredrikson_model-inversion_2015} which utilizes the gradient of a model to reconstruct its inputs.

The best possible defense against inference attacks is achieved by applying well-defined cryptographic tools, such as secure multi-party computation (SMC) \cite{ryffel_smpc_2018} or homomorphic encryption (HE) \cite{aono_he-in-ml_2017}, to either the data, the model, or the learning or inference process. Unfortunately, their application on deep neural networks are accompanied by a non-negligible performance overhead~\mbox{\cite{aono_he-in-ml_2017, ryffel_smpc_2018}}.

Differential Privacy \cite{dwork_differential-privacy_2016} is a scheme for privacy-preserving data release and a common basis for MemI defenses.
By applying deliberate noise to the ML training process, participation is hidden, e.g., on a per-sample basis, and therefore the MemI attack's success is impeded.
An extension of DP, commonly used in privacy-preserving learning, bounds the attacker's capability of \emph{distinguishability} by $\varepsilon$ with a relaxation that exceptions to the rule may occur with a likelihood of $\delta$ \cite{dwork_odo_2006}.

DP-SGD \cite{abadi_dpsgd_2016} is a variant of the standard stochastic gradient descent algorithm (on which many ML optimizer base) that applies $(\varepsilon,\delta)$-DP noise to the gradient.
PATE, by \citeauth{papernot_pate_2017}, is another exemplary approach that accomplishes DP, in which a local ensemble of private teacher models, each trained on a distinct partition of a sensitive dataset, labels a public dataset. This single-party training procedure counters the MemI attack by hiding the contribution of the teachers' labeling votes with targeted DP noise. Subsequently, the securely labeled data can be used to train a public student model.

Especially in sharing approaches, that include non-trivial training procedures or exchange of information through data or model sharing, privacy deserves great attention.
In this work, we focus primarily on utility benefit, yet we also comment on privacy in sharing approaches that we deem beneficial for our use case.

%% file: content/evaluation_setup.tex
% !TEX root = ../paper.tex

\section{Evaluation Setup}
\label{sec:evaluation_setup}
In this section, we provide an overview of our evaluation setup, including the state-of-the-art classifiers selected, the data sources used, our dataset generation scheme, the collaborative ML approaches examined, and the evaluation methodology used.

\subsection{State-of-the-Art Classifiers}
\label{sec:classifiers}

In the following, we introduce three state-of-the-art classifiers for DGA detection that will be used as part of our evaluation.
All three classifiers are based on neural networks, operate on single domain names, and output a probability that indicates whether the input domains are benign or DGA-generated.
We choose classifiers based on the following three aspects: (1) they achieve state-of-the-art detection performance, (2) they operate context-less, and (3) they are based on different types of neural network architectures.
The latter allows us to examine whether the type of architecture used has an impact on the success of collaborative ML and, secondly, to compare the potential increase in performance.
In detail, our classifier selection includes a RNN-based, a CNN-based, and a ResNet-based classifier.
In addition to the output generated, all three classifiers share the same input pre-processing steps.
Before a domain name is input to any classifier it is encoded using integer encoding (i.e., each unique character is replaced by a unique integer).
The encoded domain names are then consumed by an embedding layer in order to incorporate semantics into the encoding.
From then on, each classifier processes the embedded domain names differently, which we briefly present below.
Detailed information on the implementations of the individual classifiers can be found in \cite{woodbridge_predicting_2016, yu_character_2018, drichel_analyzing_2020}.

\paragraph{\textbf{Endgame}}
Woodbridge et al.~\cite{woodbridge_predicting_2016} proposed a RNN classifier based on a long short-term memory (LSTM) layer consisting of 128 nodes with sigmoid activation. We refer to this classifier as Endgame in the following.

\paragraph{\textbf{NYU}}
Yu et al.~\cite{yu_character_2018} proposed a CNN-based model that uses two stacked 1-dimensional convolutional layers with 128 filters for separating benign from DGA-generated domain names. We refer to this model as NYU in the following.

\paragraph{\textbf{ResNet}}
Drichel et al.~\cite{drichel_analyzing_2020} proposed a ResNet-based model for DGA binary classification. The model is build of a single residual block which incorporates a skip connection between convolutional layers. The skip connection allows the gradient to bypass the convolutional layers unaltered during training in order to counteract the vanishing gradient problem. Similar to the NYU model, the ResNet model includes two 1-dimensional convolutional layers with 128 filters in the residual.

\subsection{Data Sources}
In the following, we introduce five different data sources from which we obtain NXDs for our collaborative ML experiments.
We use one source to obtain malicious labeled samples and four distinct sources from different networks for benign labeled data.
This rich data enables us to conduct collaborative ML experiments that are similar to a real-world setting.
Moreover, the different benign data sources allow us to investigate whether the collaboratively trained classifiers generalize to different networks.

\subsubsection{Malicious Data: DGArchive}

We obtain malicious labeled domain names from the OSINT feed of DGArchive~\cite{plohmann_comprehensive_2016}. For our evaluation we include all available samples up to September 1\textsuperscript{st}, 2020. In total, DGArchive provides us with approximately 126 million unique domain names generated by 95 different DGA families.

\subsubsection{Benign Data}

We obtain benign labeled samples from four different data sources: two university networks (\textit{University\textsubscript{A}} and \textit{University\textsubscript{B}}), the networks of an \textit{Association} of universities, and the networks of a large \textit{Company}.
University\textsubscript{B} is also a member of the Association and thus the samples generated in the network of University\textsubscript{B} are also observable in the networks of the Association.
Since the time intervals of the recordings for these two data sources overlap, we remove the intersection of all samples from both records in order to prevent an artificial increase in classification performance for a jointly trained classifier if samples from both sources are used.
Otherwise, it would be possible that a classifier is evaluated on a test set that contains samples that were also used to train the classifier.
Additionally, we compare all domain names obtained from all benign data sources against DGArchive and remove all matches in order to clean our data as much as possible.
In the following, we provide a brief overview of the benign data sources.

\paragraph{\textbf{University\textsubscript{A}}}
We obtained a one-month recording of September 2019 from the central DNS resolver of RWTH Aachen University which is located in Germany. This recording comprises approximately 26 million unique NXDs that originate from academic and administrative networks, student residences' networks, and networks of the university hospital of RWTH Aachen.

\paragraph{\textbf{University\textsubscript{B}}}
We obtained a one-month recording from mid-May 2020 until mid-June 2020 from the networks of Masaryk University which is located in the Czech Republic. This recording contains approximately 8 million unique samples.

\paragraph{\textbf{Association}}
We received additional benign samples from CESNET: an association of universities of the Czech Republic and the Czech Academy of Sciences consisting of 27 members in total.
CESNET operates and develops the national e-infrastructure for science, research, and education.
From this data source, we obtained a subset of occurred NXDs from the day recording of 2020-06-15.
In total, we obtained approximately 362k unique samples.

\paragraph{\textbf{Company}}
We obtained a one-month recording of July 2019 that comprises approximately 21 million unique NXDs from several DNS resolvers of Siemens AG which is a large company that operates in Asia, Europe, and in the USA.

\subsection{Dataset Generation}
\label{sec:dataset_generation}
\begin{figure}[!t]
	\centering
	\includegraphics[width=0.573\columnwidth]{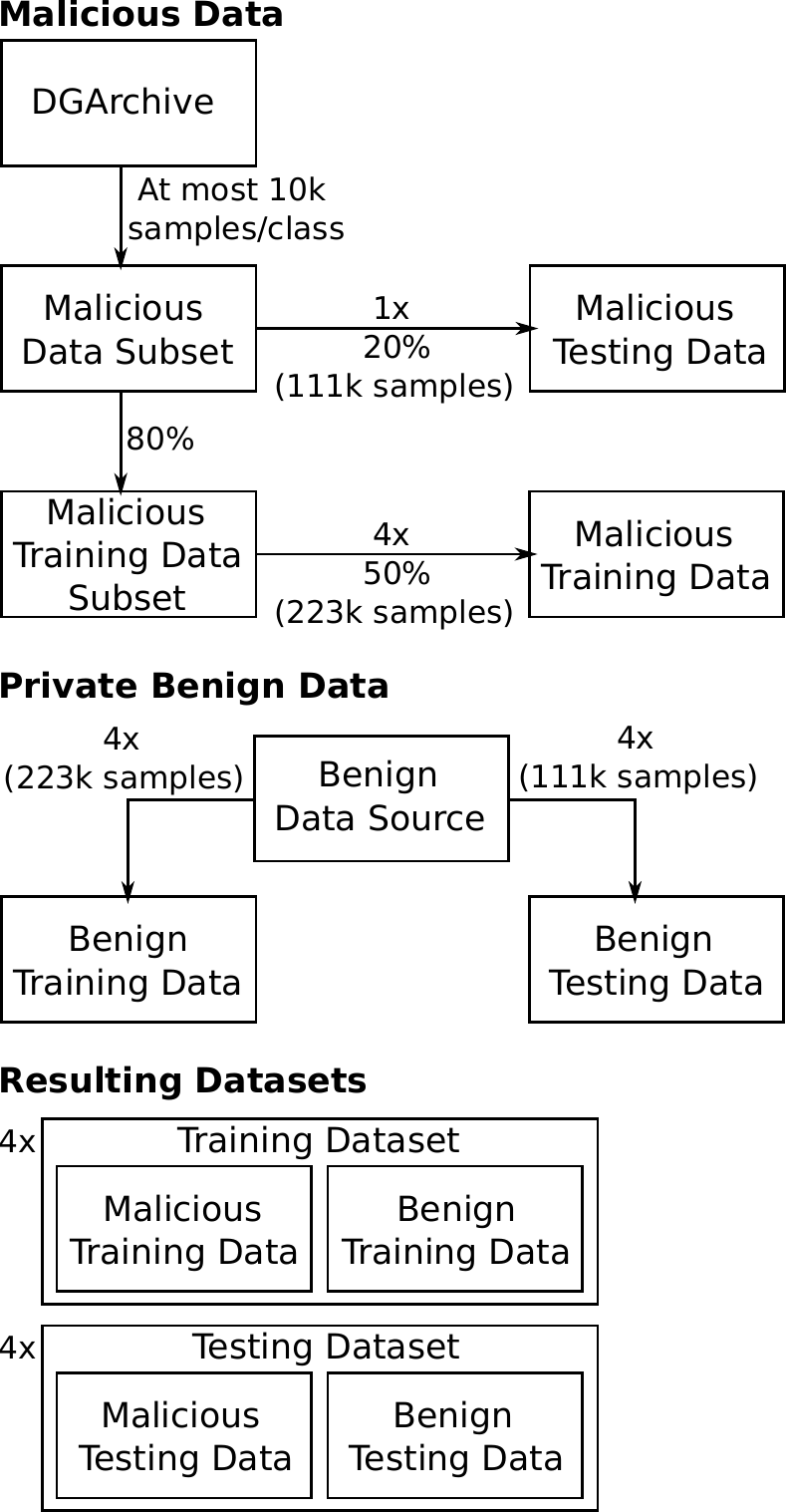}
	\caption{Datasets  Generation Scheme}
	\label{fig:datasets}
\end{figure}

In the following, we first describe our process of generating suitable datasets for our experiments using the above data sources.
We provide an illustration of this process in Fig.~\ref{fig:datasets} for convenience.

In order to create diverse datasets and to cope with the large amount of available training samples, we first subsample our malicious labeled data into a smaller set that includes at most 10k samples per DGA family.
We include all samples for DGA families for which less than 10k samples are available.
Thereby, we also include samples from underrepresented DGA families.
We do this because in ~\cite{drichel_making_2020} it was shown that by including a few samples to the training of a classifier, its detection performance for underrepresented DGAs can be increased significantly without reducing its detection rates for well-represented DGAs.

From the selected subset we then split 20\% (approximately 111k samples) stratified across all included DGA families for the test sets.
For each of our benign data sources we select individual malicious training data by subsampling 50\% (approximately 223k samples) from the remaining malicious labeled samples.
By subsampling from a larger common pool of malicious samples, we can create four training sets that contain both duplicate and unique malicious samples.
We do this because, in a real-world scenario, it is very likely that the collaborating parties are using overlapping sets of malicious samples.
Note, in contrast to the benign labeled samples, the malicious samples are available in public repositories and are not privacy-sensitive.

The benign samples are not shared between different training sets as they are considered to be the main privacy-critical aspect of the collaborative DGA detection use case.
We carry out a similar selection process for the benign training and testing samples.
From each of the four benign data sources, we randomly subsample the same amount of benign training and testing samples as we did for malicious training and testing samples, respectively.

We use these data selections to create four training and testing dataset pairs, one for each benign data source.
To this end, we combine the respective malicious and benign data selections to balanced training and testing datasets.
Note that during this dataset generation we ensured that the samples included in the training and testing datasets are completely disjoint.
Each of the four training and testing datasets include approximately 446k and 223k samples, respectively.

Additionally, we create two balanced training datasets that include publicly available benign data using the same generation process.
These datasets are used to train initial global models for our FL experiments.
The public benign data originates from the Tranco list~\cite{lepochat_tranco_2019}, which contains a ranking of the most popular domains that has been hardened against manipulation.
Using this data we create two datasets, one contains the top entries of the list, while the benign data of the other dataset consists of random samples.

\subsection{Methodology \& Sharing Approaches}
Using these datasets, we are able to precisely measure the influence of collaborative ML on the classification performance of various classifiers.
To obtain meaningful results, we repeat the whole dataset generation process five times and thereby create 20 individual training and testing dataset pairs which include malicious labeled samples from DGArchive and benign data from the four benign data sources.
By this means, we also generate ten training datasets used in the FL experiments for training an initial global model using publicly available data.
In the following, we repeat every experiment five times and present the averages of the individual results.
Note, the datasets are generated similarly to a five-fold cross validation, i.e., the testing datasets are completely disjoint with both the training datasets and the testing datasets within the repetitions.

In this work, we train all classifiers using early stopping with a patience of three epochs to avoid overfitting and assess their performance during training on holdout sets that consist of  random 5\% splits of the used training data.

To measure the impact of collaborative ML on classification performance, we evaluate all possible combinations of participants for each examined approach.
Overall, our comprehensive study consists of 13,440 evaluation passes.
The total number of individual experiments differs between the various collaborative ML approaches.

In the following, we provide an overview on the sharing approaches we consider, all of which address the problem of collaborative ML.
We first present the baseline and then the different sharing approaches including the number of experiments per approach.

\subsubsection{Baseline}
All sharing approaches are compared against the baseline to evaluate their performance.
The baseline evaluations are similar to traditional training and testing of a classifier using training data from a single organization.
Each organization trains their own model using their own private benign data and malicious training samples from DGArchive.
No training data is shared among any organizations and also no global model is derived.

We train one classifier for each of the four organizations and evaluate them on every available test dataset.
To this end, we perform five repetitions of training and testing for four organizations (University\textsubscript{A}, University\textsubscript{B}, Association, Company) and three classifier models (Endgame, NYU, ResNet), thus perform $5\cdot4\cdot4\cdot3=240$ baseline evaluations in total.

\subsubsection{Ensemble Classification}
In Ensemble classification a global classifier is build using the classifiers trained by each organization.
Similar to the baseline scenario each organization first trains a classifier using their own private benign data.
These classifiers are then shared with all participants.
Each party now combines the individual classifiers to an Ensemble classifier.
The combination of the classifiers can be done (1) by using a majority voting system on the binary labels (hard labels) or (2) by averaging the results of the individual classifiers to a single confidence score (soft labels).
A tie is possible when using the majority voting system with an even number of participants.
In such a case, we resort to the soft labels approach, where we average all predictions and predict a domain name as malicious if the average is greater than 0.5 and benign otherwise.

In total there are $\binom{4}{2} + \binom{4}{3} + \binom{4}{4} = 11$ possible combinations of organizations for building a combined model using four different parties.
Here, we also evaluate classifiers on all four different testing datasets, regardless of the combination of local classifiers used.
Thereby, we are able to measure the generalization capability of classifiers on benign data from unknown networks.
In total, we perform five repetitions using eleven possible organization combinations, two ensemble approaches (hard and soft labels), four test datasets, and three classifier models ($5\cdot11\cdot2\cdot4\cdot3=1320$ evaluation passes).

\subsubsection{Federated Learning}
Federated learning (FL)~\cite{mcmahan_communication-efficient_2017} is a technique to train a classifier collaboratively without sharing local data.
First a global model is initialized that is shared among all participants in the collaborative training.
This global model can either be randomly initialized using standard initialization methods or can be pre-trained using non-sensitive public data.
In this work, we evaluate FL using three different initial global models, one that is randomly initialized (Random Model), and two pre-trained models.
For pre-training a global model, we make use of malicious samples from DGArchive and benign domain names from the Tranco list~\cite{lepochat_tranco_2019}.
This list contains a ranking of the most popular domain names, which is also protected against manipulation.
One pre-trained model uses the top entries of the Tranco list for benign domain names, while the other model uses random samples.
In the following, we refer to these models as \textit{Tranco Top} and \textit{Tranco Random}, respectively.
After the global model is shared among all participants an iterative training procedure is performed.
The global model is trained locally by each organization using their own private training data for each federation step.
The weight updates to the global model in each federation step are then shared with all other participants, such that everyone can now average the weight updates of the current step and add them to the global classifiers weights.
Thereby, each party obtains the same global model that is then used within the next federation step.
This iterative training process continues until the global model converges.
The only data shared between the organizations are the model weight updates after each federation step and the initial global model.
In this work, we investigate two federation approaches.
In the first approach, we federate after each local model epoch (\textit{Federation after Model Epoch}), while in the second approach we only federate once after all local models have converged (\textit{Federation after Model Convergence}).

Here, we perform five repetitions using eleven organization combinations, three initial global models (Tranco Top, Tranco Random, Random Model), two possibilities for federation (after model epoch, after model convergence), four test datasets, and three classifier models ($5\cdot11\cdot3\cdot2\cdot4\cdot3=3960$ evaluation passes).

\subsubsection{Feature Extractor Sharing}
This sharing approach is related to Transfer Learning.
All classifiers under consideration (Section~\ref{sec:classifiers}) are based on DL models that make use of a fully connected (dense) layer to output the final classification score.
This layer can be viewed as a sort of classification layer that performs a logistic regression for binary classification.
The output of this layer is a confidence score that indicates whether an input domain is benign (score $< 0.5$) or malicious (score $\ge 0.5$).
All layers before this classification layer can be treated as a feature extractor, which produces features used for classification by the final output layer.
Instead of sharing the complete classifier as in the naive Ensemble approach, here, we hope to reduce the model's privacy leakage by sharing less layers.
Thus in this approach, each organization trains a model based on their own private training data.
Subsequently, the trained feature extractors are shared among all participants.
Each organization now combines their own and received feature extractors to a new model.
To this end, the feature extractors are applied in parallel and their outputs are concatenated and flattened.
Additionally, a new dense classification layer is appended to the new model.
This classification layer is not trained yet, thus the organizations freeze the weights of the feature extractors and use local training data to train the classification layer separately.
In the end, each organization obtains a model which incorporates information about samples occurring in the other organizations through the shared feature extractors.
The resulting models are not identical, since the training of the classification layer is performed on private training samples.

For this approach, each organization first trains a classifier using its own private benign data and derives an individual feature extractor.
Then we combine the four feature extractors into eleven possible classifiers. 
In contrast to the two above mentioned approaches, here we require additional training to fit the final classification layer that combines the results of the shared feature extractors.
Hence, in total we perform five repetitions using eleven possible organization combinations with four training datasets, four test datasets, and three classifier models ($5\cdot11\cdot4\cdot4\cdot3=2640$ evaluation passes).

\subsubsection{Teacher-Student}
The last examined sharing approach is based on a Teacher-Student (T/S) setup.
Here, an organization queries trained classifiers of other organizations (teacher) in order to obtain labels for their own data.
This labeled data is then used by the querying organization to train an own classifier (student).
Using this approach the teacher classifiers are not exposed to the organization that is training the student classifier.
Thereby, white-box attacks against the privacy of an organization that provides the labeling service are not applicable.
Usually more than one teacher is involved in the labeling process of a training sample, thus the individual labels or scores need to be combined.
Similar to Ensemble classification, we examine two approaches: (1) majority voting on binary labels (hard labels) and (2) soft labeling by averaging confidence scores.
When using the majority voting approach, a tie is resolved in the same way as in Ensemble classification.
For this approach no global shared model is trained, instead every party again derives an individual model similar  to Feature Extractor Sharing.

We train classifiers that are similar to the baseline classifiers as teacher models for this approach.
Similar to Feature Extractor Sharing, here we need a training dataset that is labeled by the teachers and used for training a student classifier.
Thus, in total we perform five repetitions using eleven possible organization combinations with four training datasets, two teacher result combination approaches (hard and soft labeling), four test datasets, and three classifier models ($5\cdot11\cdot4\cdot2\cdot4\cdot3=5280$ evaluation passes).

\subsubsection{Evaluation Metrics}
\label{sec:metrics}
As the benign training samples are identified as the main privacy-critical aspect of this use case we mainly use the false positive rate (FPR) as a proxy to determine the possible gain or loss in classification performance.
Moreover, we argue that a low FPR is the most important attribute of a suitable classifier.
Otherwise normal operation of a network would not be possible with too many false alarms.
For the sake of completeness, we additionally provide the results for the accuracy (ACC) and the true positive rate (TPR) which is a proxy for determining the amount of detected DGA-generated domain names.
Note that we use the same malicious samples in all four testing datasets within an experiment repetition.
Thereby, we reduce the impact of the malicious samples on the ACC metric and thus ease the measurement of the classifiers' generalization ability for unseen benign samples that come from different networks.

\subsubsection{Sharing Scenarios}
\label{sec:sharing_scenarios}
The sharing approaches are evaluated in different scenarios which are derived from research questions on possible real-world application environments for trained classifiers.

\paragraph{\textbf{Best Case}}
In this scenario, multiple network operators jointly train a classifier and are mostly interested in a good performance in their own networks.
This is related to the following research question: is collaborative training beneficial for organizations that mostly classify data from their own distribution? 
In this best-case scenario, we provide averaged results for classifiers that are evaluated only on the test datasets containing samples coming from the organizations involved in the training.

\paragraph{\textbf{Average Case}}
The average of all evaluations is used as a general performance indicator of the trained classifiers for each collaborative ML approach.
We use this scenario for a comparative evaluation of the different sharing approaches.

\paragraph{\textbf{Worst Case}}
The worst-case scenario is contrasting the best-case scenario.
Here, classifiers are evaluated on all test datasets that contain samples from organizations that have not participated in the classifier training. 
Using this scenario, we examine the generalization capability of collaboratively trained classifiers (i.e., whether the classifiers improve in their detection performance for samples originating from different networks).

%% file: content/evaluation.tex
% !TEX root = ../paper.tex
\section{Evaluation Results}
\label{sec:evaluation}
In this section, we present the results of our comprehensive study.
First, we highlight differences between the four organizations' data and provide an overview of the performance in  the three sharing scenarios.
Subsequently, we present the results of our comparative evaluation.
Finally, we analyze the effect of the number of participants in collaborative ML.

\subsection{Network Differences \& Sharing Scenarios}
To better explain the actual evaluation steps and to detail the calculations done for the different sharing scenarios we present the average scores for five repetitions of the baseline experiment using the Endgame classifier in Table~\ref{tab:endgame_baseline_detailed}. We provide the results for the Endgame classifier as an example. The results for the NYU and ResNet models are similar.

\input{tables/endgame_baseline_detailed.tex}

Here we list the average scores for the five repetitions of Endgame classifiers per training dataset used and per test dataset separately. 

The TPRs per training network are equal for all test networks as we use the same malicious samples within all four test datasets within an repetition (see Section~\ref{sec:dataset_generation} and Section~\ref{sec:metrics}).

The best FPRs on the individual test networks are always achieved by the classifiers that were trained using benign samples which originate from the same network as the testing samples.
This is expected since those classifiers are specifically trained to extract and classify characteristics of the benign domain names from the respective network.
For example, benign samples from different networks may miss certain features or exhibit other characteristics.
The average of the table entries where the train network is equal to the test network represents the best-case scenario and is presented at the bottom of the table.

The classifiers trained using benign samples from distinct networks achieve different results for the individual test datasets.
Samples from the Association are most commonly classified wrong.
In some cases the FPR for these evaluations is even greater than 6\%. 
We reckon that this is due to the fact that the samples from this network are the most diverse as they originate from over 27 different organizations.
Moreover, we filtered out the intersection of the samples from the University\textsubscript{B} network with the samples from the Association as both networks are interconnected.
Thereby, we likely removed easily recognizable samples that are naturally occurring in both networks.
This could be the reason for the larger FPRs for University\textsubscript{B} and the Association compared to those for the other two networks.

Similarly to the best case, we provide the results for the average and worst case in the lower part of Table~\ref{tab:endgame_baseline_detailed}.
While the average case is calculated using the average of all table entries, the worst case only contains the results of the entries for which the train network differs from the test network.
The results for the different sharing scenarios are not of interest for the baseline evaluation.
As could be expect, the best case results are better than the average case results, which are better than the worst case results.

\subsection{Sharing Approaches}
\label{sec:sharing_approaches}
In this section, we compare the different approaches for collaborative ML.
First, for comparison and to show that training on publicly available data is not enough for DGA detection on private data, we display the averaged results achieved by the two different types of pre-trained models that we use within our FL experiments for all three classifier types over all test datasets in Table~\ref{tab:public_data_eval}.

\input{tables/public_data_eval.tex}

All six trained classifiers yield  high FPRs between 33\% and 59\%, indicating that training on publicly available data alone is insufficient for classifying privacy-sensitive domain names.

In Table~\ref{tab:avg_comparison}, we present the results for the average case, for all classifiers, and all collaborative ML approaches examined.

\input{tables/avg_comparison.tex}

In order to assess whether the collaborative training is beneficial we additionally provide the baseline results at the top of the table.
For convenience, we color table entries red if the scores are worse than the ones of the baseline and green otherwise.
All approaches except for the FL setting \textit{Random Model - Model Convergence} achieve better TPRs than the baseline.
This is an expected outcome because the training datasets used by the individual organizations contain additional malicious labeled samples from which a collaboratively trained classifier can learn.
Thus, due to collaboration, intelligence about additional malicious labeled training samples is combined in the jointly trained classifiers.

The only exception is the FL setting \textit{Random Model - Model Convergence}.
In this setting, we use a randomly initialized model and federate the updates of the local models after they converged.
While the NYU and the ResNet model are still functional and only achieve slightly worse classification scores, the TPR of the Endgame classifier falls from over 99.9\% (baseline) to 55\%.
We reckon the reason to be that the model updates from a random initialized model to a fully converged model vary quite large and the individual organizations optimize their models to different local optima.
Averaging and applying all model updates may result in an non-optimal global model.
The Endgame model is far more affected by this compared to the CNN-based NYU and ResNet model.
This is due to the fact that RNNs are processing inputs sequentially.
Averaging the weight updates of fully converged models that are used to process sequential data can thus result in a non-functional global model.
In the following, we exclude the FL setting \textit{Random Model - Model Convergence} from our study and mainly focus on the FPR for our assessment.

All other FL scenarios lead to an improvement compared to the baseline results.
Here, the Endgame model performs significantly better in scenarios where a pre-trained initial global model was used.
We assume that this is due to the fact that when using a pre-trained model, there are significantly fewer gradients towards local optima for participants to optimize their models.
Similar to the FL setting \textit{Random Model - Model Convergence}, this is an important property, especially for RNN-based classifiers. 
In contrast, the ResNet model achieves the best results using the randomly initialized model.
In all FL settings, federating after each model epoch achieves better results than federating after model convergence.

The only other approach that leads to better classification results is Feature Extractor Sharing. 
The achieved results are comparable to the ones achieved by FL.

Ensemble classification and the T/S approach lead to worse results than the baseline.
Comparing both approaches, the T/S approach yields a lower FPR for all three classifier types.
Furthermore, with either approach, it makes little difference whether soft or hard labels are used.

While the absolute improvement achieved by collaborative ML may seem rather small, the relative reduction in the FPR is significant and could be decisive for the real-world application of classifiers.
Compared to the baseline classifiers, the best approaches achieve on average a FPR reduction of 51.7\%, 31.9\%, and 44.3\% for Endgame, NYU, and ResNet, respectively.

In summary, additional malicious samples in collaborative ML improve the TPRs for all sharing approaches.
In the average-case scenario, only the collaborative ML approaches Feature Extractor Sharing and FL are advantageous for the use case of DGA detection.
Using federation after model epoch leads to better results than federation after model convergence in FL.

\subsection{Collaboration Analysis}
In this section, our goal is to determine whether an increasing number of participants has a positive or negative effect on the classification performance of jointly trained classifiers.
To this end, we investigate two scenarios.

In the first scenario, we make use of the best-case scenario defined in Section~\ref{sec:sharing_scenarios}.
Here, organizations want to use jointly trained classifiers to classify samples from their own networks most of the time.
Thus, our goal is to determine whether the classification performance on those samples increases or decreases with an increasing number of participants. 
Thereby, organizations can decide whether or not it makes sense to use a collaboratively trained classifier for their own network.

In the second scenario, the worst case, we want to determine whether an increasing collaboration improves the generalization capabilities of the classifiers and thus the classification performance on samples from external networks.

We again use the FPR as a proxy to determine the performance of the classifiers.
In Table~\ref{tab:fpr_best_worst_case}, we present the achieved FPR scores for the different collaborative ML approaches and classifier types separated by the number of participants.

\input{tables/fpr_best_worst_case.tex}

For visibility, we omit the ACC and the TPR metric, however, most of the time a better or worse FPR correlates with a better or worse ACC.
In this evaluation we are not primarily interested in comparing the achieved scores of the different approaches with the performance of the baseline that is presented at the top of the table.
Rather, we are interested in whether an increasing cooperation improves or worsens the performance achieved.
Thus, we color code the entries different to Table~\ref{tab:avg_comparison}.
Here, we mark all table entries green if they are always improving with an increasing number of participants.
When the approaches produce consecutive worse results we color them red.
We do not color any entries for approaches for which the increases or decreases in classification performance are not consecutive.
In the following, we present the evaluation results for the best and worst case in detail.

\subsubsection{Best Case}
Most of the collaborative approaches achieve (1) worse results compared to the baseline and (2) are decreasing in classification performance with increasing participants.
This behavior can be explained by the fact that the baseline's best-case scenario is the ideal training and classification setting.
There, classifiers are assessed on data that comes from the same distribution as the samples used for training.
No information of samples from other organizations are incorporated in those classifiers.
Thereby, the baseline classifiers are specialized in classifying samples that originate from the same network as the training samples used.
Thus, it is not surprising that the baseline classifiers achieve almost the best results compared to the other approaches.
The collaborative ML approaches, on the other hand, incorporate also information of samples from other networks.
Thereby, they are less specialized in classifying samples from a single network but rather are more generalized and thus achieve worse results compared to the baseline.
The fact that these approaches achieve worse results with an increasing number of participants can be explained similarly.
The more participants, the less the classifiers are specialized on samples of a single network.
%For example, the same trained baseline classifiers are used in Ensemble classification.
For example, Ensemble classification uses classifiers that are similar to the baseline classifiers. 
However, the more classifiers there are in an ensemble, the less influence a single classifier has on the final classification result. 
Therefore, in the best-case scenario, the classification performance decreases.

The only approach that improves with an increasing number of participants is Feature Extractor Sharing.
Moreover, for all three classifier types, Feature Extractor Sharing achieves better FPRs than the baseline.
This is because this approach creates models that are similar to the baseline but also incorporates additional information about samples from other networks via feature extractors that are applied in parallel.
Since the shared feature extractors are not retrained, information about samples from individual networks is very well preserved with this approach. 
In addition, the intelligence incorporated in the shared feature extractors is harnessed by this approach and leads to improvement even beyond baseline. 
Note, although the differences in FPRs are rather small, we argue that our results are significant because of the large amount of evaluations done and the fact that this behavior is observable for all three types of classifiers.

From these results it can be seen that only Feature Extractor Sharing is beneficial in the best-case scenario, where organizations want to use collaborative ML classifiers to classify samples from their own network most of the time.

\subsubsection{Worst Case}
In this scenario, we evaluate whether an increasing number of participants improves the detection performance of jointly trained classifiers for samples that originate from external networks.
Since we only have four different sources of benign data, the maximum number of participants in this scenario is three. 

The results obtained for the Ensemble classification deteriorate as the number of participants increases for all three classifiers. 
The FPRs for the T/S approach improve only for the Endgame classifier.
However, the achieved rates are worse than those of the baseline.

For FL, the FPRs improve in all settings and for all classifiers except for Endgame when a randomly initialized model is used.
We assume that this is due to the same reasons given in Section~\ref{sec:sharing_approaches}.
The ResNet classifier, however, achieves the best results using a randomly initialized model.
The achieved FPRs for Endgame and ResNet using federation after model epoch are significantly lower than for the approaches that make use of federation after model convergence.
For the NYU classifier, no significant difference can be measured for the various models.

For Feature Extractor Sharing, the FPRs achieved improve with an increased number of participants for all three classifier types.
However, the rates are significantly worse than the ones achieved by Endgame and ResNet using FL with federation after model epoch.
For NYU, the rates are comparable to those seen in the FL settings.

In summary, in the worst-case scenario, only the Feature Extractor Sharing and FL approaches improve in performance with an increasing number of participants and achieve better scores than the baseline.
FL with federation after model epoch achieves best results for Endgame and ResNet and thus generalizes best to different networks.
When comparing the different types of classifiers, Endgame and ResNet are better suited for FL than NYU.
For RNN-based classifiers pre-trained initial models should be used.

%% file: tables/endgame_baseline_detailed.tex
% !TEX root = ../paper.tex

\begin{table}[!t]
	\caption{Averaged Baseline Results for Endgame Classifier}
	\label{tab:endgame_baseline_detailed}
	\centering
		\resizebox{\columnwidth}{!}{
	\begin{tabular}{lcccc}
		\toprule
		\textbf{Train Network} & \textbf{Test Network} & \textbf{ACC} & \textbf{TPR} & \textbf{FPR} \\
		\midrule
		\multirow{4}{*}{\textbf{University\textsubscript{A}}} & \textbf{University\textsubscript{A}} & 0.99851 & 0.99968 & 0.00267 \\
		& University\textsubscript{B} 											   					 & 0.99304 & 0.99968 & 0.01359 \\
		& Association     					  											   			 & 0.96892 & 0.99968 & 0.06184 \\
		& Company 					  											   					 & 0.99834 & 0.99968 & 0.00300 \\
		\midrule
		\multirow{4}{*}{\textbf{University\textsubscript{B}}} & University\textsubscript{A} 		 & 0.99717 & 0.99966 & 0.00532 \\
		& \textbf{University\textsubscript{B}} 											    		 & 0.99808 & 0.99966 & 0.00350 \\
		& Association 						  											   			 & 0.97180 & 0.99966 & 0.05606 \\
		& Company 					  											   					 & 0.99853 & 0.99966 & 0.00259 \\
		\midrule
		\multirow{4}{*}{\textbf{Association}} & 	University\textsubscript{A}						 & 0.99674 & 0.99739 & 0.00390 \\
		& 	University\textsubscript{B}	    				  					   					 & 0.99444 & 0.99739 & 0.00852 \\
		& 	\textbf{Association} 											  					     & 0.99645 & 0.99739 & 0.00450 \\
		& 	Company											  					   					 & 0.99771 & 0.99739 & 0.00196 \\
		\midrule
		\multirow{4}{*}{\textbf{Company}} & University\textsubscript{A}					    		 & 0.98923 & 0.99968 & 0.02122 \\
		& University\textsubscript{B}					  						   					 & 0.99037 & 0.99968 & 0.01894 \\
		& Association											  						   			 & 0.96833 & 0.99968 & 0.06302 \\
		& \textbf{Company}										  						    		 & 0.99888 & 0.99968 & 0.00192 \\
		\midrule
		\midrule
		\multicolumn{2}{c}{\textbf{Best Case}} &  0.99798  & 0.99910 & 0.00315 \\
		\multicolumn{2}{c}{\textbf{Average Case}} &  0.99103  & 0.99910 & 0.01703 \\
		\multicolumn{2}{c}{\textbf{Worst Case}} &  0.98872  & 0.99910 & 0.02166 \\
		\bottomrule
	\end{tabular}
		}
\end{table}

%% file: tables/public_data_eval.tex
% !TEX root = ../paper.tex

\begin{table}[!t]
	\caption{Results of Pre-trained Models using Public Data}
	\label{tab:public_data_eval}
	\centering
%	\resizebox{\columnwidth}{!}{
		\begin{tabular}{lcccc}
			\toprule
			\textbf{Classifier} & \textbf{Benign Data} & \textbf{ACC} & \textbf{TPR} & \textbf{FPR} \\
			\midrule
			\multirow{2}{*}{Endgame} & Tranco Top & 0.68329 & 0.95492 & 0.58834 \\
			& Tranco Random & 0.67730 & 0.95054 & 0.59594 \\
			\midrule
			\multirow{2}{*}{NYU} & Tranco Top & 0.68612 & 0.94876 & 0.57652 \\
			& Tranco Random & 0.71336 & 0.94310 & 0.51637 \\
			\midrule
			\multirow{2}{*}{ResNet} & Tranco Top & 0.78667 & 0.94952 & 0.37617 \\
			& Tranco Random & 0.79899 & 0.93619 & 0.33821 \\
			\bottomrule
		\end{tabular}
%	}
\end{table}

%% file: tables/avg_comparison.tex
% !TEX root = ../paper.tex
\begin{table*}[!t]
	\caption{Results of the Average Case Including All Classifier Types, and All Collaborative ML Approaches}
	\label{tab:avg_comparison}
	\centering
	\resizebox{\linewidth}{!}{
		\begin{tabular}{l|ccc|ccc|ccc}
%			\toprule
			\hline
			\multirow{2.5}{*}{\textbf{Approach}} & \multicolumn{3}{c|}{\textbf{Endgame}} &  \multicolumn{3}{c|}{\textbf{NYU}} &  \multicolumn{3}{c}{\hspace{-2mm}\textbf{ResNet}} \Tstrut\\
			\cline{2-4}
			\cline{5-7} 
			\cline{8-10}
			& \textbf{ACC} & \textbf{TPR} & \textbf{FPR}  &  \textbf{ACC} & \textbf{TPR} & \textbf{FPR} & \textbf{ACC} & \textbf{TPR} & \hspace{0.1mm} \textbf{FPR} \Tstrut \\
			\hline
			\textbf{Baseline} 							& \hspace{0.1mm} 0.99103 & \hspace{0.1mm} 0.99910 & \hspace{0.1mm} 0.01703 & \hspace{0.1mm} 0.99151 & \hspace{0.1mm} 0.99903 & \hspace{0.1mm} 0.01600 & \hspace{0.1mm} 0.99102 & \hspace{0.1mm} 0.99844 &  \hspace{0.1mm} 0.01640 \Tstrut\\ % baseline
			\hline
			\textbf{Ensemble:} Soft Labels 							& \cellcolor{red!50}   0.98812 & \cellcolor{green!50} 0.99975 & \cellcolor{red!50}   0.02352 & \cellcolor{red!50}   0.98870 & \cellcolor{green!50} 0.99976 & \cellcolor{red!50}   0.02236 & \cellcolor{red!50}   0.98834 & \cellcolor{green!50} 0.99946 & \cellcolor{red!50}  0.02279 \Tstrut\\ % ensemble_soft_labels
			\textbf{Ensemble:} Hard Labels  							& \cellcolor{red!50}   0.98842 & \cellcolor{green!50} 0.99981 & \cellcolor{red!50}   0.02296 & \cellcolor{red!50}   0.98901 & \cellcolor{green!50} 0.99976 & \cellcolor{red!50}   0.02174 & \cellcolor{red!50}   0.98926 & \cellcolor{green!50} 0.99909 & \cellcolor{red!50} \hspace{-2mm}  0.02057 \\ % ensemble_majority_vote
			\hline
			\textbf{FL:} Random Model - Model Convergence  & \cellcolor{red!50}   0.76755 & \cellcolor{red!50}    0.55131 & \cellcolor{green!50}  0.01621 & \cellcolor{red!50}   0.98300 & \cellcolor{red!50}	 0.98764 & \cellcolor{red!50}   0.02163 & \cellcolor{red!50}   0.98671 & \cellcolor{red!50}   0.99278 & \cellcolor{red!50}  0.01937 \Tstrut\\ % federated_learning_null_model_converged
			\textbf{FL:} Random Model - Model Epoch        & \cellcolor{green!50} 0.99396 & \cellcolor{green!50}  0.99968 & \cellcolor{green!50}  0.01177 & \cellcolor{green!50} 0.99392 & \cellcolor{green!50} 0.99983 & \cellcolor{green!50} 0.01199 & \cellcolor{green!50} 0.99511 & \cellcolor{green!50} 0.99935 & \cellcolor{green!50} \hspace{-2mm} 0.00913 \\ % federated_learning_null_model_epoch
			\textbf{FL:} Tranco Top - Model Convergence    & \cellcolor{green!50} 0.99336 & \cellcolor{green!50}  0.99958 & \cellcolor{green!50}  0.01286 & \cellcolor{green!50} 0.99325 & \cellcolor{green!50} 0.99975 & \cellcolor{green!50} 0.01326 & \cellcolor{green!50} 0.99291 & \cellcolor{green!50} 0.99974 & \cellcolor{green!50} \hspace{-2mm} 0.01393 \\ % federated_learning_tranco_top_model_converged
			\textbf{FL:} Tranco Top - Model Epoch          & \cellcolor{green!50} 0.99553 & \cellcolor{green!50}  0.99956 & \cellcolor{green!50}  0.00850 & \cellcolor{green!50} 0.99400 & \cellcolor{green!50} 0.99981 & \cellcolor{green!50} 0.01181 & \cellcolor{green!50} 0.99491 & \cellcolor{green!50} 0.99922 & \cellcolor{green!50} \hspace{-2mm} 0.00940 \\ % federated_learning_tranco_top_model_epoch
			\textbf{FL:} Tranco Random - Model Convergence & \cellcolor{green!50} 0.99365 & \cellcolor{green!50}  0.99946 & \cellcolor{green!50}  0.01216 & \cellcolor{green!50} 0.99360 & \cellcolor{green!50} 0.99978 & \cellcolor{green!50} 0.01257 & \cellcolor{green!50} 0.99268 & \cellcolor{green!50} 0.99967 & \cellcolor{green!50} \hspace{-2mm} 0.01431 \\ % federated_learning_tranco_random_model_converged
			\textbf{FL:} Tranco Random - Model Epoch       & \cellcolor{green!50} 0.99565 & \cellcolor{green!50}  0.99952 & \cellcolor{green!50}  0.00823 & \cellcolor{green!50} 0.99413 & \cellcolor{green!50} 0.99978 & \cellcolor{green!50} 0.01153 & \cellcolor{green!50} 0.99498 & \cellcolor{green!50} 0.99929 & \cellcolor{green!50} \hspace{-2mm} 0.00934 \\ % federated_learning_tranco_random_model_epoch
			\hline
			\textbf{Feature Extractor Sharing}  		& \cellcolor{green!50} 0.99394 & \cellcolor{green!50}   0.99921 & \cellcolor{green!50} 0.01133 & \cellcolor{green!50} 0.99411 & \cellcolor{green!50} 0.99913 & \cellcolor{green!50} 0.01090 & \cellcolor{green!50} 0.99307 & \cellcolor{green!50} 0.99880 & \cellcolor{green!50} 0.01266 \Tstrut\\ % first_n_layers
			\hline				
			\textbf{T/S:} Soft Labels  							& \cellcolor{red!50}   0.98979 & \cellcolor{green!50} 0.99963 & \cellcolor{red!50}   0.02006 & \cellcolor{red!50}   0.99029 & \cellcolor{green!50} 0.99965 & \cellcolor{red!50}   0.01908 & \cellcolor{red!50}   0.99001 & \cellcolor{green!50} 0.99949 & \cellcolor{red!50}  0.01948 \Tstrut\\ % teacher_student_soft_labels_labeled
			\textbf{T/S:} Hard Labels  							& \cellcolor{red!50}   0.98966 & \cellcolor{green!50} 0.99968 & \cellcolor{red!50}   0.02036 & \cellcolor{red!50}   0.99017 & \cellcolor{green!50} 0.99965 & \cellcolor{red!50}   0.01930 & \cellcolor{red!50}   0.99026 & \cellcolor{green!50} 0.99950 & \cellcolor{red!50}   \hspace{-2mm} 0.01898 \\ % teacher_student_majority_vote_labeled
%			\bottomrule
			\hline
		\end{tabular}
	}
\end{table*}

%% file: tables/fpr_best_worst_case.tex
% !TEX root = ../paper.tex

\begin{table*}[!t]
	\caption{FPRs of the Best and Worst Case, Separated by Number of Participants}
	\label{tab:fpr_best_worst_case}
	\centering
%	\resizebox{\linewidth}{!}{
		\begin{tabular}{l|c|ccc|ccc}
			\hline
			\multirow{2.5}{*}{\textbf{Approach}} & \multirow{2.5}{*}{\textbf{Parties}} & \multicolumn{3}{c|}{\textbf{Best Case}} &  \multicolumn{3}{c}{\textbf{Worst Case}}  \Tstrut\\
			\cline{3-5}
			\cline{6-8}
			& & \textbf{Endgame} & \textbf{NYU} & \textbf{ResNet}  & \textbf{Endgame} & \textbf{NYU} & \textbf{ResNet}  \Tstrut\\
			\hline
			\textbf{Baseline} 							& -   & \hspace{0.1mm} 0.00315 & \hspace{0.1mm} 0.00328 & \hspace{0.1mm} 0.00326  & \hspace{0.1mm} 0.02166  & \hspace{0.1mm} 0.02024  & \hspace{0.1mm} 0.02078 \Tstrut \\ % baseline
			\hline
			  											& 2   & \cellcolor{red!50}   0.02149 & \cellcolor{red!50}   0.02023 & \cellcolor{red!50}   0.02081  & \cellcolor{red!50}   0.02420  & \cellcolor{red!50}   0.02306  & \cellcolor{red!50}   0.02376 \Tstrut\\ % ensemble_soft_labels
			\textbf{Ensemble:} Soft Labels				& 3   & \cellcolor{red!50}   0.02393 & \cellcolor{red!50}   0.02266 & \cellcolor{red!50}   0.02316  & \cellcolor{red!50}   0.02493  & \cellcolor{red!50}   0.02409  & \cellcolor{red!50}   \hspace{-2mm} 0.02387 \\ % ensemble_soft_labels
														& 4   & \cellcolor{red!50}   0.02489 & \cellcolor{red!50}   0.02402 & \cellcolor{red!50}   0.02366  & -        & -        & -       \\ % ensemble_soft_labels
			\hline
														& 2   & \cellcolor{red!50}   0.02068 & \cellcolor{red!50}   0.01978 & \cellcolor{red!50}   0.01887  & \cellcolor{red!50}   0.02399  & \cellcolor{red!50}   0.02242  & \cellcolor{red!50}   0.02127 \Tstrut\\ % ensemble_majority_vote
			\textbf{Ensemble:} Hard Labels				& 3   & \cellcolor{red!50}   0.02332 & \cellcolor{red!50}   0.02207 & \cellcolor{red!50}   0.02077  & \cellcolor{red!50}   0.02462  & \cellcolor{red!50}   0.02322  & \cellcolor{red!50}   \hspace{-2mm} 0.02186 \\ % ensemble_majority_vote
														& 4   & \cellcolor{red!50}   0.02404 & \cellcolor{red!50}   0.02316 & \cellcolor{red!50}   0.02167  & -        & -        & -       \\ % ensemble_majority_vote
			\hline
			                                                & 2 & \cellcolor{red!50} 0.00464 & \cellcolor{red!50} 0.00620 & \cellcolor{red!50} 0.00465 & \cellcolor{red!50}   0.02032 & \cellcolor{green!50} 0.01886 & \cellcolor{green!50} 0.01668  \Tstrut\\ % federated_learning_null_model_epoch		2
			\textbf{FL:} Random Model - Model Epoch         & 3 & \cellcolor{red!50} 0.00738 & \cellcolor{red!50} 0.00922 & \cellcolor{red!50} 0.00516 & \cellcolor{red!50}   0.02122 & \cellcolor{green!50} 0.01786 & \cellcolor{green!50} \hspace{-2mm} 0.01473  \\ % federated_learning_null_model_epoch		3
															& 4 & \cellcolor{red!50} 0.01120 & \cellcolor{red!50} 0.01120 & \cellcolor{red!50} 0.00624 & -        & -        & -       \\ \hline % federated_learning_null_model_epoch		4
															& 2 & \cellcolor{red!50} 0.00803 & \cellcolor{red!50} 0.00783 & \cellcolor{red!50} 0.00856 & \cellcolor{green!50} 0.01769 & \cellcolor{green!50} 0.01841 & \cellcolor{green!50} 0.01924  \Tstrut\\ % federated_learning_tranco_top_model_converged		2
			\textbf{FL:}  Tranco Top - Model Convergence    & 3 & \cellcolor{red!50} 0.01147 & \cellcolor{red!50} 0.01170 & \cellcolor{red!50} 0.01230 & \cellcolor{green!50} 0.01723 & \cellcolor{green!50} 0.01834 & \cellcolor{green!50} \hspace{-2mm} 0.01877  \\ % federated_learning_tranco_top_model_converged		3
															& 4 & \cellcolor{red!50} 0.01267 & \cellcolor{red!50} 0.01365 & \cellcolor{red!50} 0.01415 & -        & -        & -       \\ \hline % federated_learning_tranco_top_model_converged		4
															& 2 & \cellcolor{red!50} 0.00361 & \cellcolor{red!50} 0.00535 & \cellcolor{red!50} 0.00410 & \cellcolor{green!50} 0.01571 & \cellcolor{green!50} 0.01905 & \cellcolor{green!50} 0.01655  \Tstrut\\ % federated_learning_tranco_top_model_epoch		2
			\textbf{FL:}  Tranco Top - Model Epoch          & 3 & \cellcolor{red!50} 0.00491 & \cellcolor{red!50} 0.00879 & \cellcolor{red!50} 0.00597 & \cellcolor{green!50} 0.01492 & \cellcolor{green!50} 0.01830 & \cellcolor{green!50} \hspace{-2mm} 0.01573  \\ % federated_learning_tranco_top_model_epoch		3
															& 4 & \cellcolor{red!50} 0.00591 & \cellcolor{red!50} 0.01205 & \cellcolor{red!50} 0.00780 & -        & -        & -       \\ \hline % federated_learning_tranco_top_model_epoch		4
															& 2 & \cellcolor{red!50} 0.00719 & \cellcolor{red!50} 0.00745 & \cellcolor{red!50} 0.00974 & \cellcolor{green!50} 0.01763 & \cellcolor{green!50} 0.01799 & \cellcolor{green!50} 0.01889 \Tstrut \\ % federated_learning_tranco_random_model_converged		2
			\textbf{FL:}  Tranco Random - Model Convergence & 3 & \cellcolor{red!50} 0.01022 & \cellcolor{red!50} 0.01062 & \cellcolor{red!50} 0.01297 & \cellcolor{green!50} 0.01680 & \cellcolor{green!50} 0.01773 & \cellcolor{green!50} \hspace{-2mm} 0.01831  \\ % federated_learning_tranco_random_model_converged		3
															& 4 & \cellcolor{red!50} 0.01185 & \cellcolor{red!50} 0.01235 & \cellcolor{red!50} 0.01428 & -        & -        & -       \\ \hline % federated_learning_tranco_random_model_converged		4
															& 2 & \cellcolor{red!50} 0.00351 & \cellcolor{red!50} 0.00520 & \cellcolor{red!50} 0.00394 & \cellcolor{green!50} 0.01551 & \cellcolor{green!50} 0.01857 & \cellcolor{green!50} 0.01593  \Tstrut\\ % federated_learning_tranco_random_model_epoch		2
			\textbf{FL:}  Tranco Random - Model Epoch       & 3 & \cellcolor{red!50} 0.00458 & \cellcolor{red!50} 0.00877 & \cellcolor{red!50} 0.00599 & \cellcolor{green!50} 0.01397 & \cellcolor{green!50} 0.01775 & \cellcolor{green!50} \hspace{-2mm} 0.01557  \\ % federated_learning_tranco_random_model_epoch		3
															& 4 & \cellcolor{red!50} 0.00575 & \cellcolor{red!50} 0.01146 & \cellcolor{red!50} 0.00958 & -        & -        & -       \\ \hline % federated_learning_tranco_random_model_epoch		4
			
														& 2   & \cellcolor{green!50} 0.00306 & \cellcolor{green!50} 0.00303 & \cellcolor{green!50} 0.00314  & \cellcolor{green!50} 0.01871  & \cellcolor{green!50} 0.01756  & \cellcolor{green!50}   0.01804 \Tstrut \\ % first_n_layers
			\textbf{Feature Extractor Sharing}			& 3   & \cellcolor{green!50} 0.00295 & \cellcolor{green!50} 0.00302 & \cellcolor{green!50} 0.00307  & \cellcolor{green!50} 0.01796  & \cellcolor{green!50} 0.01722  & \cellcolor{green!50}  \hspace{-2mm} 0.01784 \\ % first_n_layers
														& 4   & \cellcolor{green!50} 0.00293 & \cellcolor{green!50} 0.00298 & \cellcolor{green!50} 0.00302  & -        & -        & -       \\ % first_n_layers
			\hline
														& 2   & \cellcolor{red!50}   0.00328 & \cellcolor{red!50}   0.00339 & \cellcolor{red!50}   0.00339  & \cellcolor{green!50} 0.02264  & \cellcolor{red!50}   0.02143  & \cellcolor{red!50}   0.02198 \Tstrut\\ % teacher_student_soft_labels_labeled
			\textbf{T/S:} Soft Labels					& 3   & \cellcolor{red!50}   0.01735 & \cellcolor{red!50}   0.01703 & \cellcolor{red!50}   0.01646  & \cellcolor{green!50} 0.02252  & \cellcolor{red!50}   0.02171  & \cellcolor{red!50}   \hspace{-2mm} 0.02247 \\ % teacher_student_soft_labels_labeled
														& 4   & \cellcolor{red!50}   0.01822 & \cellcolor{red!50}   0.01776 & \cellcolor{red!50}   0.01744  & -        & -        & -       \\ % teacher_student_soft_labels_labeled
			\hline
														& 2   & \cellcolor{red!50}   0.00329 & \hspace{0.1mm} 0.00331 & \cellcolor{red!50}   0.00333  & \cellcolor{green!50} 0.02386  & \cellcolor{red!50}   0.02132  & \cellcolor{red!50}   0.02103 \Tstrut\\ % teacher_student_majority_vote_labeled
			\textbf{T/S:} Hard Labels					& 3   & \cellcolor{red!50}   0.01725 & \hspace{0.1mm} 0.01795 & \cellcolor{red!50}   0.01650  & \cellcolor{green!50} 0.02208  & \cellcolor{red!50}   0.02246  & \cellcolor{red!50}   \hspace{-2mm} 0.02218 \\ % teacher_student_majority_vote_labeled
														& 4   & \cellcolor{red!50}   0.01766 & \hspace{0.1mm} 0.01761 & \cellcolor{red!50}   0.01694  & -        & -        & -       \\ % teacher_student_majority_vote_labeled
			\hline
		\end{tabular}
%	}
\end{table*}

%% file: content/privacy_discussion.tex
 % !TEX root = ../paper.tex
\section{Privacy Discussion}
\label{sec:privacy_discussion}

We complement our most beneficial approaches, Feature Extractor Sharing and FL, with a thorough discussion on relevant privacy-threatening inference attacks.

\subsection{Feature Extractor Sharing}

Partial models are shared in this approach, giving an adversary white-box access to the gradient computation and the weights, that are directly influenced by the data in the learning process.

\subsubsection{Model Inversion}
In a white-box setting, gradient-based Model Inversion attacks \cite{fredrikson_warfarin_2014, fredrikson_model-inversion_2015} may be deployed to iteratively reconstruct an input to the shared model utilizing its gradient and a loss comparing the desired output and the model's output for the reconstructed input.
The reason we consider this a minor threat is two-fold:
(1) Model Inversion attacks do not perform well in practice, especially if the targeted model is highly complex (in number of layers or weights) \cite{hitaj_dm-gan_2017}.
(2) Reconstructing true inputs, requires to know a set of the targeted party's feature vectors, which are however \emph{not} shared in this approach.

\subsubsection{Membership Inference}
The classic MemI attack is defined for complete classifier models, i.e., with a final softmax output. Theoretically, the MemI attack setup (as in \cite{shokri_membership-inference_2017}) could be redefined for the case of a partial model. To the best of our knowledge such research has, however, not been conducted yet.

\subsection{Federated Learning}
The FL paradigm was proposed to enhance the data privacy for participants by reducing the exposure of their data. This alone is not sufficient, as attacks still threaten data privacy in FL \cite{lyu_privacy_2020, shen_distributed_2020}.

\subsubsection{Gradient Leakage}
To retain valuable contribution from parties in FL, it is crucial to ensure the non-disclosure of a participant's local data.
However, retaining data locality in FL is not sufficient, as \cite{zhu_deep_2019, orekondy_gradient-leaks_2020, zhu_r-gap_2021} demonstrate the necessity to shield gradient updates from inspection by the aggregating instance(s) that may reconstruct or infer sensitive data.
Inference attacks during execution of the FL protocol can be rendered infeasible via a secure aggregation protocol \cite{bonawitz_practical_2017}, which computes the global average gradient in a SMC protocol that completely obstructs the inspection of local updates, thereby providing the best possible privacy.
Multiple improvements have been proposed (e.g., \cite{guo_secure_2020, so_turbo-aggregate_2021}): With the currently best performance overhead of $O(N\text{log}N)$ per FL round \cite{so_turbo-aggregate_2021}, deploying secure aggregation is easily applicable in our use case with $N\leq4$ parties.

\subsubsection{Membership Inference}
In FL, a MemI attack against the globally trained model threatens disclosure of participation on both sample level (i.e., the classic MemI attack \cite{shokri_membership-inference_2017}) as well as on client level \cite{wang_beyond_2018}.
A multitude of DP-driven research in FL exists (e.g., \cite{hitaj_dm-gan_2017, geyer_differentially_2017, zheng_federated_2021, kerkouche_constrained_2021, kim_federated_2021, kairouz_fl-advances_2021}), that propose or investigate DP-based defenses and on occasion examine the inherent privacy/utility trade-off.
Improvements of or alternatives to the classic DP-SGD for the FL setting are proposed in \cite{kairouz_practical_2021, ye_sample-based_2021}, any of which can be applied to the local training of each party.
DP yields sound privacy guarantees w.r.t. the bounds $(\varepsilon, \delta)$, the quality of which are however influenced by the individual use case and its available data and hence, application of DP would require a further assessment of the resulting privacy/utility trade-off.

\subsubsection{Byzantine Attackers}
Our work only considers the presence of trusted parties, yet for completeness we also give a quick view on Byzantine parties in FL \cite{blanchard_byzantine-tolerant-gradient-descent_2017}, that are defined by arbitrary or faulty behavior, including intentional misbehavior such as privacy-threatening Free-riding \cite{fraboni_free-rider_2021} or sabotage (as in Poisoning \cite{tolpegin_data_2020} or Backdooring \cite{sun_can_2019, bagdasaryan_how_2020}).

Distributed learning in the Byzantine setting has been studied in \cite{blanchard_byzantine-tolerant-gradient-descent_2017, malecki_simeon_2021} (and larger literature bodies referenced in \cite{so_byzantine-resilient_2020, guerraoui_differential_2021}).
\citeauth{blanchard_byzantine-tolerant-gradient-descent_2017} argue that a single Byzantine user can influence any linear aggregation mechanism, and therefore also model conversion, to an arbitrarily large extent and present their first Byzantine-tolerant defense termed Krum.
Other effective defenses have been proposed, which base on Krum or utilize similar insights that updates from malicious FL parties are separable from benign ones and can be filtered out \cite{blanchard_byzantine-tolerant-gradient-descent_2017, tolpegin_data_2020, yadav_clustering_2021, so_byzantine-resilient_2020}.
\citeauth{malecki_simeon_2021} propose a Byzantine-robust and Sybil-resistant defense.

\citeauth{so_byzantine-resilient_2020} additionally present an integration of their defense mechanism into the secure aggregation protocol by \citeauth{bonawitz_practical_2017} utilizing secure distance computation via HE.
Unfortunately, \citeauth{guerraoui_differential_2021} provide first insights into the incompatibility of DP-based and Byzantine defenses.

%% file: content/conclusion.tex
% !TEX root = ../paper.tex
\section{Conclusion}
\label{sec:conclusion}

In this paper, we performed a comprehensive study of collaborative ML for the real-world use case of DGA detection and discussed the privacy implications caused by beneficial sharing approaches.
Thereby, we identified advantageous and disadvantageous approaches for different types of classifiers and showed that collaborative ML can lead to a reduction in FPR by up to 51.7\%.
Additionally, we showed that the usage of publicly available data is insufficient for DGA detection on private data.
This shows the need for privacy-preserving collaborative ML approaches to make use of the advantages provided by classifying NXDs for DGA detection.
In two real-world cases, we have shown that greater participation in collaborative ML does not always lead to better classification results. 
In fact, we only assess  Feature Extractor Sharing and FL of the four examined collaborative ML approaches as beneficial for DGA detection. 
Feature Extractor Sharing should be used if a party wants to classify samples that come from their own network most of the time.
FL on the other hand generalizes best to unknown networks.
The four examined collaborative ML variants based on T/S learning and Ensemble classification lead to worse results than the baseline.

In terms of privacy, we have additionally discussed the applicability of inference attacks in our two beneficial approaches.
Related state-of-the-art defense mechanisms enable enhanced privacy for sharing in our use case with negligible overhead.
For the MemI attack it remains to assess the privacy/utility trade-off, that is inherent to DP-based defenses and the data in our use case.
Also, MemI on partial models (Feature Extractor Sharing) is not regarded in prior work so far.

%% file: paper.bbl

\begin{thebibliography}{65}

%%% the Web address.

\ifx \showCODEN    \undefined \def \showCODEN     #1{\unskip}     \fi
\ifx \showDOI      \undefined \def \showDOI       #1{#1}\fi
\ifx \showISBNx    \undefined \def \showISBNx     #1{\unskip}     \fi
\ifx \showISBNxiii \undefined \def \showISBNxiii  #1{\unskip}     \fi
\ifx \showISSN     \undefined \def \showISSN      #1{\unskip}     \fi
\ifx \showLCCN     \undefined \def \showLCCN      #1{\unskip}     \fi
\ifx \shownote     \undefined \def \shownote      #1{#1}          \fi
\ifx \showarticletitle \undefined \def \showarticletitle #1{#1}   \fi
\ifx \showURL      \undefined \def \showURL       {\relax}        \fi
% invisible to TeX
\providecommand\bibfield[2]{#2}
\providecommand\bibinfo[2]{#2}
\providecommand\natexlab[1]{#1}
\providecommand\showeprint[2][]{arXiv:#2}

\bibitem[\protect\citeauthoryear{Abadi, Chu, Goodfellow, McMahan, Mironov,
  Talwar, and Zhang}{Abadi et~al\mbox{.}}{2016}]%
        {abadi_dpsgd_2016}
\bibfield{author}{\bibinfo{person}{Martin Abadi}, \bibinfo{person}{Andy Chu},
  \bibinfo{person}{Ian Goodfellow}, \bibinfo{person}{H.~Brendan McMahan},
  \bibinfo{person}{Ilya Mironov}, \bibinfo{person}{Kunal Talwar}, {and}
  \bibinfo{person}{Li Zhang}.} \bibinfo{year}{2016}\natexlab{}.
\newblock \showarticletitle{Deep Learning with Differential Privacy}. In
  \bibinfo{booktitle}{\emph{Computer and Communications Security}}.
  \bibinfo{publisher}{ACM}.
\newblock


\bibitem[\protect\citeauthoryear{Al-Rubaie and Chang}{Al-Rubaie and
  Chang}{2019}]%
        {al_rubaie_ppml_2019}
\bibfield{author}{\bibinfo{person}{Mohammad Al-Rubaie} {and}
  \bibinfo{person}{J.~Morris Chang}.} \bibinfo{year}{2019}\natexlab{}.
\newblock \showarticletitle{Privacy-Preserving Machine Learning: Threats and
  Solutions}. In \bibinfo{booktitle}{\emph{Security \& Privacy}}.
  \bibinfo{publisher}{IEEE}.
\newblock


\bibitem[\protect\citeauthoryear{Antonakakis, Perdisci, Nadji, Vasiloglou,
  {Abu-Nimeh}, Lee, and Dagon}{Antonakakis et~al\mbox{.}}{2012}]%
        {antonakakis_throwaway_2012}
\bibfield{author}{\bibinfo{person}{Manos Antonakakis}, \bibinfo{person}{Roberto
  Perdisci}, \bibinfo{person}{Yacin Nadji}, \bibinfo{person}{Nikolaos
  Vasiloglou}, \bibinfo{person}{Saeed {Abu-Nimeh}}, \bibinfo{person}{Wenke
  Lee}, {and} \bibinfo{person}{David Dagon}.} \bibinfo{year}{2012}\natexlab{}.
\newblock \showarticletitle{From Throw-Away Traffic to Bots: Detecting the Rise
  of {DGA}-Based Malware}. In \bibinfo{booktitle}{\emph{{{USENIX Security
  Symposium}}}}.
\newblock
\showISBNx{978-931971-95-9}


\bibitem[\protect\citeauthoryear{Ateniese, Mancini, Spognardi, Villani, Vitali,
  and Felici}{Ateniese et~al\mbox{.}}{2015}]%
        {ateniese_hacking_2015}
\bibfield{author}{\bibinfo{person}{Giuseppe Ateniese}, \bibinfo{person}{Luigi~V
  Mancini}, \bibinfo{person}{Angelo Spognardi}, \bibinfo{person}{Antonio
  Villani}, \bibinfo{person}{Domenico Vitali}, {and} \bibinfo{person}{Giovanni
  Felici}.} \bibinfo{year}{2015}\natexlab{}.
\newblock \showarticletitle{Hacking smart machines with smarter ones: How to
  extract meaningful data from machine learning classifiers}. In
  \bibinfo{booktitle}{\emph{International Journal of Security and Networks}}.
  \bibinfo{publisher}{Inderscience Publishers}.
\newblock


\bibitem[\protect\citeauthoryear{Bagdasaryan, Veit, Hua, Estrin, and
  Shmatikov}{Bagdasaryan et~al\mbox{.}}{2020}]%
        {bagdasaryan_how_2020}
\bibfield{author}{\bibinfo{person}{Eugene Bagdasaryan},
  \bibinfo{person}{Andreas Veit}, \bibinfo{person}{Yiqing Hua},
  \bibinfo{person}{Deborah Estrin}, {and} \bibinfo{person}{Vitaly Shmatikov}.}
  \bibinfo{year}{2020}\natexlab{}.
\newblock \showarticletitle{How to Backdoor Federated Learning}. In
  \bibinfo{booktitle}{\emph{Artificial Intelligence and Statistics}}. PMLR.
\newblock


\bibitem[\protect\citeauthoryear{Bilge, Sen, Balzarotti, Kirda, and
  Kruegel}{Bilge et~al\mbox{.}}{2014}]%
        {bilge_exposure_2014}
\bibfield{author}{\bibinfo{person}{Leyla Bilge}, \bibinfo{person}{Sevil Sen},
  \bibinfo{person}{Davide Balzarotti}, \bibinfo{person}{Engin Kirda}, {and}
  \bibinfo{person}{Christopher Kruegel}.} \bibinfo{year}{2014}\natexlab{}.
\newblock \showarticletitle{Exposure: A Passive {DNS} Analysis Service to
  Detect and Report Malicious Domains}. In
  \bibinfo{booktitle}{\emph{Transactions on Information and System Security}}.
  \bibinfo{publisher}{ACM}.
\newblock


\bibitem[\protect\citeauthoryear{Blanchard, El~Mhamdi, Guerraoui, and
  Stainer}{Blanchard et~al\mbox{.}}{2017}]%
        {blanchard_byzantine-tolerant-gradient-descent_2017}
\bibfield{author}{\bibinfo{person}{Peva Blanchard}, \bibinfo{person}{El~Mahdi
  El~Mhamdi}, \bibinfo{person}{Rachid Guerraoui}, {and} \bibinfo{person}{Julien
  Stainer}.} \bibinfo{year}{2017}\natexlab{}.
\newblock \showarticletitle{Machine Learning with Adversaries: Byzantine
  Tolerant Gradient Descent}. In \bibinfo{booktitle}{\emph{{Neural}
  {Information} {Processing} {Systems}}}. \bibinfo{publisher}{Curran Associates
  Inc.}
\newblock


\bibitem[\protect\citeauthoryear{Bonawitz, Ivanov, Kreuter, Marcedone, McMahan,
  Patel, Ramage, Segal, and Seth}{Bonawitz et~al\mbox{.}}{2017}]%
        {bonawitz_practical_2017}
\bibfield{author}{\bibinfo{person}{Keith Bonawitz}, \bibinfo{person}{Vladimir
  Ivanov}, \bibinfo{person}{Ben Kreuter}, \bibinfo{person}{Antonio Marcedone},
  \bibinfo{person}{H.~Brendan McMahan}, \bibinfo{person}{Sarvar Patel},
  \bibinfo{person}{Daniel Ramage}, \bibinfo{person}{Aaron Segal}, {and}
  \bibinfo{person}{Karn Seth}.} \bibinfo{year}{2017}\natexlab{}.
\newblock \showarticletitle{Practical Secure Aggregation for Privacy-Preserving
  Machine Learning}. \bibinfo{publisher}{ACM}.
\newblock


\bibitem[\protect\citeauthoryear{Dietterich}{Dietterich}{2000}]%
        {dietterich_ensemble_2000}
\bibfield{author}{\bibinfo{person}{Thomas~G. Dietterich}.}
  \bibinfo{year}{2000}\natexlab{}.
\newblock \showarticletitle{Ensemble Methods in Machine Learning}. In
  \bibinfo{booktitle}{\emph{Multiple Classifier Systems}}.
  \bibinfo{publisher}{Springer}.
\newblock


\bibitem[\protect\citeauthoryear{Drichel, Meyer, Sch\"uppen, and
  Teubert}{Drichel et~al\mbox{.}}{2020a}]%
        {drichel_analyzing_2020}
\bibfield{author}{\bibinfo{person}{Arthur Drichel}, \bibinfo{person}{Ulrike
  Meyer}, \bibinfo{person}{Samuel Sch\"uppen}, {and} \bibinfo{person}{Dominik
  Teubert}.} \bibinfo{year}{2020}\natexlab{a}.
\newblock \showarticletitle{Analyzing the Real-World Applicability of {DGA}
  Classifiers}. In \bibinfo{booktitle}{\emph{Conference on Availability,
  Reliability and Security}}. \bibinfo{publisher}{ACM}.
\newblock


\bibitem[\protect\citeauthoryear{Drichel, Meyer, Sch\"uppen, and
  Teubert}{Drichel et~al\mbox{.}}{2020b}]%
        {drichel_making_2020}
\bibfield{author}{\bibinfo{person}{Arthur Drichel}, \bibinfo{person}{Ulrike
  Meyer}, \bibinfo{person}{Samuel Sch\"uppen}, {and} \bibinfo{person}{Dominik
  Teubert}.} \bibinfo{year}{2020}\natexlab{b}.
\newblock \showarticletitle{Making Use of {NXt} to Nothing: Effect of Class
  Imbalances on {DGA} Detection Classifiers}. In
  \bibinfo{booktitle}{\emph{Conference on Availability, Reliability and
  Security}}. \bibinfo{publisher}{ACM}.
\newblock


\bibitem[\protect\citeauthoryear{Dwork}{Dwork}{2006}]%
        {dwork_differential-privacy_2016}
\bibfield{author}{\bibinfo{person}{Cynthia Dwork}.}
  \bibinfo{year}{2006}\natexlab{}.
\newblock \showarticletitle{Differential Privacy}. In
  \bibinfo{booktitle}{\emph{Automata, Languages and Programming}}.
  \bibinfo{publisher}{Springer}.
\newblock


\bibitem[\protect\citeauthoryear{Dwork, Kenthapadi, McSherry, Mironov, and
  Naor}{Dwork et~al\mbox{.}}{2006}]%
        {dwork_odo_2006}
\bibfield{author}{\bibinfo{person}{Cynthia Dwork}, \bibinfo{person}{Krishnaram
  Kenthapadi}, \bibinfo{person}{Frank McSherry}, \bibinfo{person}{Ilya
  Mironov}, {and} \bibinfo{person}{Moni Naor}.}
  \bibinfo{year}{2006}\natexlab{}.
\newblock \showarticletitle{Our Data, Ourselves: Privacy Via Distributed Noise
  Generation}. In \bibinfo{booktitle}{\emph{Advances in Cryptology}}.
  \bibinfo{publisher}{Springer}.
\newblock


\bibitem[\protect\citeauthoryear{Fraboni, Vidal, and Lorenzi}{Fraboni
  et~al\mbox{.}}{2021}]%
        {fraboni_free-rider_2021}
\bibfield{author}{\bibinfo{person}{Yann Fraboni}, \bibinfo{person}{Richard
  Vidal}, {and} \bibinfo{person}{Marco Lorenzi}.}
  \bibinfo{year}{2021}\natexlab{}.
\newblock \showarticletitle{Free-rider Attacks on Model Aggregation in
  Federated Learning}. In \bibinfo{booktitle}{\emph{Artificial Intelligence and
  Statistics}}. PMLR.
\newblock


\bibitem[\protect\citeauthoryear{Fredrikson, Jha, and Ristenpart}{Fredrikson
  et~al\mbox{.}}{2015}]%
        {fredrikson_model-inversion_2015}
\bibfield{author}{\bibinfo{person}{Matt Fredrikson}, \bibinfo{person}{Somesh
  Jha}, {and} \bibinfo{person}{Thomas Ristenpart}.}
  \bibinfo{year}{2015}\natexlab{}.
\newblock \showarticletitle{Model Inversion Attacks That Exploit Confidence
  Information and Basic Countermeasures}. In \bibinfo{booktitle}{\emph{Computer
  and Communications Security}}. \bibinfo{publisher}{ACM}.
\newblock


\bibitem[\protect\citeauthoryear{Fredrikson, Lantz, Jha, Lin, Page, and
  Ristenpart}{Fredrikson et~al\mbox{.}}{2014}]%
        {fredrikson_warfarin_2014}
\bibfield{author}{\bibinfo{person}{Matthew Fredrikson}, \bibinfo{person}{Eric
  Lantz}, \bibinfo{person}{Somesh Jha}, \bibinfo{person}{Simon Lin},
  \bibinfo{person}{David Page}, {and} \bibinfo{person}{Thomas Ristenpart}.}
  \bibinfo{year}{2014}\natexlab{}.
\newblock \showarticletitle{Privacy in Pharmacogenetics: An End-to-End Case
  Study of Personalized Warfarin Dosing}. In \bibinfo{booktitle}{\emph{USENIX
  Security Symposium}}.
\newblock


\bibitem[\protect\citeauthoryear{Geyer, Klein, and Nabi}{Geyer
  et~al\mbox{.}}{2017}]%
        {geyer_differentially_2017}
\bibfield{author}{\bibinfo{person}{Robin~C Geyer}, \bibinfo{person}{Tassilo
  Klein}, {and} \bibinfo{person}{Moin Nabi}.} \bibinfo{year}{2017}\natexlab{}.
\newblock \showarticletitle{Differentially Private Federated Learning: A Client
  Level Perspective}.
\newblock \bibinfo{journal}{\emph{arXiv:1712.07557}}.
\newblock


\bibitem[\protect\citeauthoryear{Goodfellow, Bengio, Courville, and
  Bengio}{Goodfellow et~al\mbox{.}}{2016}]%
        {goodfellow_deep-learning_2016}
\bibfield{author}{\bibinfo{person}{Ian Goodfellow}, \bibinfo{person}{Yoshua
  Bengio}, \bibinfo{person}{Aaron Courville}, {and} \bibinfo{person}{Yoshua
  Bengio}.} \bibinfo{year}{2016}\natexlab{}.
\newblock \bibinfo{booktitle}{\emph{Deep Learning}}.
\newblock \bibinfo{publisher}{MIT press}.
\newblock


\bibitem[\protect\citeauthoryear{Grill, Nikolaev, Valeros, and Rehak}{Grill
  et~al\mbox{.}}{2015}]%
        {grill_detecting_2015}
\bibfield{author}{\bibinfo{person}{Martin Grill}, \bibinfo{person}{Ivan
  Nikolaev}, \bibinfo{person}{Veronica Valeros}, {and} \bibinfo{person}{Martin
  Rehak}.} \bibinfo{year}{2015}\natexlab{}.
\newblock \showarticletitle{Detecting {DGA} Malware Using NetFlow}. In
  \bibinfo{booktitle}{\emph{IFIP/IEEE International Symposium on Integrated
  Network Management}}.
\newblock


\bibitem[\protect\citeauthoryear{Guerraoui, Gupta, Pinot, Rouault, and
  Stephan}{Guerraoui et~al\mbox{.}}{2021}]%
        {guerraoui_differential_2021}
\bibfield{author}{\bibinfo{person}{Rachid Guerraoui}, \bibinfo{person}{Nirupam
  Gupta}, \bibinfo{person}{Rafaël Pinot}, \bibinfo{person}{Sébastien
  Rouault}, {and} \bibinfo{person}{John Stephan}.}
  \bibinfo{year}{2021}\natexlab{}.
\newblock \showarticletitle{Differential {Privacy} and {Byzantine} {Resilience}
  in {SGD}: {Do} {They} {Add} {Up}?}
\newblock \bibinfo{journal}{\emph{arXiv:2102.08166}}.
\newblock


\bibitem[\protect\citeauthoryear{Guo, Liu, Lam, Zhao, Chen, and Xing}{Guo
  et~al\mbox{.}}{2020}]%
        {guo_secure_2020}
\bibfield{author}{\bibinfo{person}{Jiale Guo}, \bibinfo{person}{Ziyao Liu},
  \bibinfo{person}{Kwok-Yan Lam}, \bibinfo{person}{Jun Zhao},
  \bibinfo{person}{Yiqiang Chen}, {and} \bibinfo{person}{Chaoping Xing}.}
  \bibinfo{year}{2020}\natexlab{}.
\newblock \showarticletitle{Secure {Weighted} {Aggregation} in {Federated}
  {Learning}}.
\newblock \bibinfo{journal}{\emph{arXiv:2010.08730}}.
\newblock


\bibitem[\protect\citeauthoryear{Hinton, Vinyals, and Dean}{Hinton
  et~al\mbox{.}}{2015}]%
        {hinton_distilling_2015}
\bibfield{author}{\bibinfo{person}{Geoffrey Hinton}, \bibinfo{person}{Oriol
  Vinyals}, {and} \bibinfo{person}{Jeff Dean}.}
  \bibinfo{year}{2015}\natexlab{}.
\newblock \showarticletitle{Distilling the Knowledge in a Neural Network}.
\newblock \bibinfo{journal}{\emph{arXiv:1503.02531}}.
\newblock


\bibitem[\protect\citeauthoryear{Hitaj, Ateniese, and Perez-Cruz}{Hitaj
  et~al\mbox{.}}{2017}]%
        {hitaj_dm-gan_2017}
\bibfield{author}{\bibinfo{person}{Briland Hitaj}, \bibinfo{person}{Giuseppe
  Ateniese}, {and} \bibinfo{person}{Fernando Perez-Cruz}.}
  \bibinfo{year}{2017}\natexlab{}.
\newblock \showarticletitle{Deep Models Under the GAN: Information Leakage from
  Collaborative Deep Learning}. In \bibinfo{booktitle}{\emph{Computer and
  Communications Security}}. \bibinfo{publisher}{ACM}.
\newblock


\bibitem[\protect\citeauthoryear{Kairouz, McMahan, Song, Thakkar, Thakurta, and
  Xu}{Kairouz et~al\mbox{.}}{2021}]%
        {kairouz_practical_2021}
\bibfield{author}{\bibinfo{person}{Peter Kairouz}, \bibinfo{person}{Brendan
  McMahan}, \bibinfo{person}{Shuang Song}, \bibinfo{person}{Om Thakkar},
  \bibinfo{person}{Abhradeep Thakurta}, {and} \bibinfo{person}{Zheng Xu}.}
  \bibinfo{year}{2021}\natexlab{}.
\newblock \showarticletitle{Practical and {Private} ({Deep}) {Learning} without
  {Sampling} or {Shuffling}}.
\newblock \bibinfo{journal}{\emph{arXiv:2103.00039}}.
\newblock


\bibitem[\protect\citeauthoryear{Kairouz and McMahan}{Kairouz and
  McMahan}{2021}]%
        {kairouz_fl-advances_2021}
\bibfield{author}{\bibinfo{person}{Peter Kairouz} {and}
  \bibinfo{person}{H.~Brendan McMahan}.} \bibinfo{year}{2021}\natexlab{}.
\newblock \showarticletitle{Advances and Open Problems in Federated Learning}.
\newblock \bibinfo{journal}{\emph{Foundations and Trends in Machine Learning}}.
\newblock


\bibitem[\protect\citeauthoryear{Kerkouche, Ács, Castelluccia, and
  Genevès}{Kerkouche et~al\mbox{.}}{2021}]%
        {kerkouche_constrained_2021}
\bibfield{author}{\bibinfo{person}{Raouf Kerkouche}, \bibinfo{person}{Gergely
  Ács}, \bibinfo{person}{Claude Castelluccia}, {and} \bibinfo{person}{Pierre
  Genevès}.} \bibinfo{year}{2021}\natexlab{}.
\newblock \showarticletitle{Constrained {Differentially} {Private} {Federated}
  {Learning} for {Low}-bandwidth {Devices}}.
\newblock \bibinfo{journal}{\emph{arXiv:2103.00342}}.
\newblock


\bibitem[\protect\citeauthoryear{Kim, Gunlu, and Schaefer}{Kim
  et~al\mbox{.}}{2021}]%
        {kim_federated_2021}
\bibfield{author}{\bibinfo{person}{Muah Kim}, \bibinfo{person}{Onur Gunlu},
  {and} \bibinfo{person}{Rafael~F. Schaefer}.} \bibinfo{year}{2021}\natexlab{}.
\newblock \bibinfo{title}{Federated Learning with Local Differential Privacy:
  Trade-offs between Privacy, Utility, and Communication}.
\newblock \bibinfo{howpublished}{Cryptology ePrint Archive, Report 2021/142}.
\newblock


\bibitem[\protect\citeauthoryear{Le~Pochat, Van~Goethem, Tajalizadehkhoob,
  Korczynski, and Joosen}{Le~Pochat et~al\mbox{.}}{2019}]%
        {lepochat_tranco_2019}
\bibfield{author}{\bibinfo{person}{Victor Le~Pochat}, \bibinfo{person}{Tom
  Van~Goethem}, \bibinfo{person}{Samaneh Tajalizadehkhoob},
  \bibinfo{person}{Maciej Korczynski}, {and} \bibinfo{person}{Wouter Joosen}.}
  \bibinfo{year}{2019}\natexlab{}.
\newblock \showarticletitle{Tranco: A Research-Oriented Top Sites Ranking
  Hardened Against Manipulation}. In \bibinfo{booktitle}{\emph{Network and
  Distributed System Security Symposium}}. \bibinfo{publisher}{{Internet
  Society}}.
\newblock


\bibitem[\protect\citeauthoryear{Lyu, Yu, Ma, Sun, Zhao, Yang, and Yu}{Lyu
  et~al\mbox{.}}{2020}]%
        {lyu_privacy_2020}
\bibfield{author}{\bibinfo{person}{Lingjuan Lyu}, \bibinfo{person}{Han Yu},
  \bibinfo{person}{Xingjun Ma}, \bibinfo{person}{Lichao Sun},
  \bibinfo{person}{Jun Zhao}, \bibinfo{person}{Qiang Yang}, {and}
  \bibinfo{person}{Philip~S. Yu}.} \bibinfo{year}{2020}\natexlab{}.
\newblock \showarticletitle{Privacy and {Robustness} in {Federated} {Learning}:
  {Attacks} and {Defenses}}.
\newblock \bibinfo{journal}{\emph{arXiv:2012.06337}}.
\newblock


\bibitem[\protect\citeauthoryear{Malecki, Paik, Ignjatovic, Blair, and
  Bertino}{Malecki et~al\mbox{.}}{2021}]%
        {malecki_simeon_2021}
\bibfield{author}{\bibinfo{person}{Nicholas Malecki},
  \bibinfo{person}{Hye-young Paik}, \bibinfo{person}{Aleksandar Ignjatovic},
  \bibinfo{person}{Alan Blair}, {and} \bibinfo{person}{Elisa Bertino}.}
  \bibinfo{year}{2021}\natexlab{}.
\newblock \showarticletitle{Simeon -- {Secure} {Federated} {Machine} {Learning}
  {Through} {Iterative} {Filtering}}.
\newblock \bibinfo{journal}{\emph{arXiv:2103.07704}}.
\newblock


\bibitem[\protect\citeauthoryear{McMahan, Moore, Ramage, Hampson, and
  y~Arcas}{McMahan et~al\mbox{.}}{2017}]%
        {mcmahan_communication-efficient_2017}
\bibfield{author}{\bibinfo{person}{Brendan McMahan}, \bibinfo{person}{Eider
  Moore}, \bibinfo{person}{Daniel Ramage}, \bibinfo{person}{Seth Hampson},
  {and} \bibinfo{person}{Blaise~Aguera y Arcas}.}
  \bibinfo{year}{2017}\natexlab{}.
\newblock \showarticletitle{Communication-Efficient Learning of Deep Networks
  from Decentralized Data}. In \bibinfo{booktitle}{\emph{Artificial
  Intelligence and Statistics}}. PMLR.
\newblock


\bibitem[\protect\citeauthoryear{Nasr, Shokri, and Houmansadr}{Nasr
  et~al\mbox{.}}{2019}]%
        {nasr_comprehensive_2019}
\bibfield{author}{\bibinfo{person}{Milad Nasr}, \bibinfo{person}{Reza Shokri},
  {and} \bibinfo{person}{Amir Houmansadr}.} \bibinfo{year}{2019}\natexlab{}.
\newblock \showarticletitle{Comprehensive Privacy Analysis of Deep Learning:
  Passive and Active White-box Inference Attacks against Centralized and
  Federated Learning}. In \bibinfo{booktitle}{\emph{Security and Privacy}}.
  \bibinfo{publisher}{IEEE}.
\newblock


\bibitem[\protect\citeauthoryear{Orekondy, Oh, Zhang, Schiele, and
  Fritz}{Orekondy et~al\mbox{.}}{2020}]%
        {orekondy_gradient-leaks_2020}
\bibfield{author}{\bibinfo{person}{Tribhuvanesh Orekondy},
  \bibinfo{person}{Seong~Joon Oh}, \bibinfo{person}{Yang Zhang},
  \bibinfo{person}{Bernt Schiele}, {and} \bibinfo{person}{Mario Fritz}.}
  \bibinfo{year}{2020}\natexlab{}.
\newblock \showarticletitle{Gradient-{Leaks}: {Understanding} and {Controlling}
  {Deanonymization} in {Federated} {Learning}}.
\newblock \bibinfo{journal}{\emph{arXiv:1805.05838}}.
\newblock


\bibitem[\protect\citeauthoryear{Pan and Yang}{Pan and Yang}{2010}]%
        {pan_survey_2010}
\bibfield{author}{\bibinfo{person}{Sinno~Jialin Pan} {and}
  \bibinfo{person}{Qiang Yang}.} \bibinfo{year}{2010}\natexlab{}.
\newblock \showarticletitle{A {Survey} on {Transfer} {Learning}}. In
  \bibinfo{booktitle}{\emph{Transactions on Knowledge and Data Engineering}}.
  \bibinfo{publisher}{IEEE}.
\newblock


\bibitem[\protect\citeauthoryear{Papernot, Abadi, Úlfar Erlingsson,
  Goodfellow, and Talwar}{Papernot et~al\mbox{.}}{2017}]%
        {papernot_pate_2017}
\bibfield{author}{\bibinfo{person}{Nicolas Papernot}, \bibinfo{person}{Martín
  Abadi}, \bibinfo{person}{Úlfar Erlingsson}, \bibinfo{person}{Ian
  Goodfellow}, {and} \bibinfo{person}{Kunal Talwar}.}
  \bibinfo{year}{2017}\natexlab{}.
\newblock \showarticletitle{Semi-supervised Knowledge Transfer for Deep
  Learning from Private Training Data}.
\newblock \bibinfo{journal}{\emph{arXiv:2012.02670}}.
\newblock


\bibitem[\protect\citeauthoryear{Papernot, McDaniel, Sinha, and
  Wellman}{Papernot et~al\mbox{.}}{2018}]%
        {papernot_sok_2018}
\bibfield{author}{\bibinfo{person}{Nicolas Papernot}, \bibinfo{person}{Patrick
  McDaniel}, \bibinfo{person}{Arunesh Sinha}, {and} \bibinfo{person}{Michael~P.
  Wellman}.} \bibinfo{year}{2018}\natexlab{}.
\newblock \showarticletitle{SoK: Security and Privacy in Machine Learning}. In
  \bibinfo{booktitle}{\emph{European Symposium on Security and Privacy}}.
  \bibinfo{publisher}{IEEE}.
\newblock


\bibitem[\protect\citeauthoryear{Peck, Nie, Sivaguru, Grumer, Olumofin, Yu,
  Nascimento, and De~Cock}{Peck et~al\mbox{.}}{2019}]%
        {peck_charbot_2019}
\bibfield{author}{\bibinfo{person}{Jonathan Peck}, \bibinfo{person}{Claire
  Nie}, \bibinfo{person}{Raaghavi Sivaguru}, \bibinfo{person}{Charles Grumer},
  \bibinfo{person}{Femi Olumofin}, \bibinfo{person}{Bin Yu},
  \bibinfo{person}{Anderson Nascimento}, {and} \bibinfo{person}{Martine
  De~Cock}.} \bibinfo{year}{2019}\natexlab{}.
\newblock \showarticletitle{CharBot: A Simple and Effective Method for Evading
  DGA Classifiers}.
\newblock \bibinfo{journal}{\emph{IEEE Access}}  \bibinfo{volume}{7}
  (\bibinfo{year}{2019}).
\newblock


\bibitem[\protect\citeauthoryear{Phong, Aono, Hayashi, Wang, and Moriai}{Phong
  et~al\mbox{.}}{2018}]%
        {aono_he-in-ml_2017}
\bibfield{author}{\bibinfo{person}{Le~Trieu Phong}, \bibinfo{person}{Yoshinori
  Aono}, \bibinfo{person}{Takuya Hayashi}, \bibinfo{person}{Lihua Wang}, {and}
  \bibinfo{person}{Shiho Moriai}.} \bibinfo{year}{2018}\natexlab{}.
\newblock \showarticletitle{Privacy-Preserving Deep Learning via Additively
  Homomorphic Encryption}. In \bibinfo{booktitle}{\emph{Transactions on
  Information Forensics and Security}}. \bibinfo{publisher}{IEEE}.
\newblock


\bibitem[\protect\citeauthoryear{Plohmann, Yakdan, Klatt, Bader, and
  {Gerhards-Padilla}}{Plohmann et~al\mbox{.}}{2016}]%
        {plohmann_comprehensive_2016}
\bibfield{author}{\bibinfo{person}{Daniel Plohmann}, \bibinfo{person}{Khaled
  Yakdan}, \bibinfo{person}{Michael Klatt}, \bibinfo{person}{Johannes Bader},
  {and} \bibinfo{person}{Elmar {Gerhards-Padilla}}.}
  \bibinfo{year}{2016}\natexlab{}.
\newblock \showarticletitle{A Comprehensive Measurement Study of Domain
  Generating Malware}. In \bibinfo{booktitle}{\emph{USENIX Security
  Symposium}}.
\newblock
\showISBNx{978-1-931971-32-4}


\bibitem[\protect\citeauthoryear{Razavian, Azizpour, Sullivan, and
  Carlsson}{Razavian et~al\mbox{.}}{2014}]%
        {razavian_cnn-offtheshelf_2014}
\bibfield{author}{\bibinfo{person}{Ali~Sharif Razavian},
  \bibinfo{person}{Hossein Azizpour}, \bibinfo{person}{Josephine Sullivan},
  {and} \bibinfo{person}{Stefan Carlsson}.} \bibinfo{year}{2014}\natexlab{}.
\newblock \showarticletitle{CNN Features Off-the-Shelf: An Astounding Baseline
  for Recognition}. In \bibinfo{booktitle}{\emph{Computer Vision and Pattern
  Recognition Workshops}}. \bibinfo{publisher}{IEEE}.
\newblock


\bibitem[\protect\citeauthoryear{Ryffel, Trask, Dahl, Wagner, Mancuso,
  Rueckert, and Passerat-Palmbach}{Ryffel et~al\mbox{.}}{2018}]%
        {ryffel_smpc_2018}
\bibfield{author}{\bibinfo{person}{Theo Ryffel}, \bibinfo{person}{Andrew
  Trask}, \bibinfo{person}{Morten Dahl}, \bibinfo{person}{Bobby Wagner},
  \bibinfo{person}{Jason~V. Mancuso}, \bibinfo{person}{Daniel Rueckert}, {and}
  \bibinfo{person}{Jonathan Passerat-Palmbach}.}
  \bibinfo{year}{2018}\natexlab{}.
\newblock \showarticletitle{A generic framework for privacy preserving deep
  learning}.
\newblock \bibinfo{journal}{\emph{arXiv:1811.04017}}.
\newblock


\bibitem[\protect\citeauthoryear{Sagi and Rokach}{Sagi and Rokach}{2018}]%
        {sagi_ensemble_2018}
\bibfield{author}{\bibinfo{person}{Omer Sagi} {and} \bibinfo{person}{Lior
  Rokach}.} \bibinfo{year}{2018}\natexlab{}.
\newblock \showarticletitle{Ensemble learning: A survey}.
\newblock \bibinfo{journal}{\emph{Wiley Interdisciplinary Reviews: Data Mining
  and Knowledge Discovery}} \bibinfo{volume}{8}, \bibinfo{number}{4}
  (\bibinfo{year}{2018}).
\newblock


\bibitem[\protect\citeauthoryear{Saxe and Berlin}{Saxe and Berlin}{2017}]%
        {saxe_expose_2017}
\bibfield{author}{\bibinfo{person}{Joshua Saxe} {and}
  \bibinfo{person}{Konstantin Berlin}.} \bibinfo{year}{2017}\natexlab{}.
\newblock \showarticletitle{{eXpose}: A Character-Level Convolutional Neural
  Network with Embeddings For Detecting Malicious {URLs}, File Paths and
  Registry Keys}.
\newblock \bibinfo{journal}{\emph{arXiv:1702.08568}}.
\newblock


\bibitem[\protect\citeauthoryear{Schiavoni, Maggi, Cavallaro, and
  Zanero}{Schiavoni et~al\mbox{.}}{2014}]%
        {schiavoni_phoenix_2014}
\bibfield{author}{\bibinfo{person}{Stefano Schiavoni},
  \bibinfo{person}{Federico Maggi}, \bibinfo{person}{Lorenzo Cavallaro}, {and}
  \bibinfo{person}{Stefano Zanero}.} \bibinfo{year}{2014}\natexlab{}.
\newblock \showarticletitle{Phoenix: {DGA}-Based Botnet Tracking and
  Intelligence}. In \bibinfo{booktitle}{\emph{Detection of Intrusions and
  Malware, and Vulnerability Assessment}}. \bibinfo{publisher}{{Springer}}.
\newblock
\showISBNx{978-3-319-08509-8}


\bibitem[\protect\citeauthoryear{Sch\"uppen, Teubert, Herrmann, and
  Meyer}{Sch\"uppen et~al\mbox{.}}{2018}]%
        {schuppen_fanci_2018}
\bibfield{author}{\bibinfo{person}{Samuel Sch\"uppen}, \bibinfo{person}{Dominik
  Teubert}, \bibinfo{person}{Patrick Herrmann}, {and} \bibinfo{person}{Ulrike
  Meyer}.} \bibinfo{year}{2018}\natexlab{}.
\newblock \showarticletitle{{FANCI}: Feature-Based Automated NXDomain
  Classification and Intelligence}. In \bibinfo{booktitle}{\emph{{{USENIX
  Security Symposium}}}}.
\newblock
\showISBNx{978-1-931971-46-1}


\bibitem[\protect\citeauthoryear{Shen, Zhu, Wu, Wang, and Zhou}{Shen
  et~al\mbox{.}}{2020}]%
        {shen_distributed_2020}
\bibfield{author}{\bibinfo{person}{Sheng Shen}, \bibinfo{person}{Tianqing Zhu},
  \bibinfo{person}{Di Wu}, \bibinfo{person}{Wei Wang}, {and}
  \bibinfo{person}{Wanlei Zhou}.} \bibinfo{year}{2020}\natexlab{}.
\newblock \showarticletitle{From distributed machine learning to federated
  learning: In the view of data privacy and security}.
\newblock \bibinfo{journal}{\emph{Concurrency and Computation: Practice and
  Experience}}.
\newblock


\bibitem[\protect\citeauthoryear{Shi, Chen, and Li}{Shi et~al\mbox{.}}{2018}]%
        {shi_malicious_2018}
\bibfield{author}{\bibinfo{person}{Yong Shi}, \bibinfo{person}{Gong Chen},
  {and} \bibinfo{person}{Juntao Li}.} \bibinfo{year}{2018}\natexlab{}.
\newblock \showarticletitle{Malicious Domain Name Detection Based on Extreme
  Machine Learning}.
\newblock \bibinfo{journal}{\emph{Neural Processing Letters}}
  \bibinfo{volume}{48}, \bibinfo{number}{3}.
\newblock
\showISSN{1573-773X}


\bibitem[\protect\citeauthoryear{Shokri, Stronati, Song, and Shmatikov}{Shokri
  et~al\mbox{.}}{2017}]%
        {shokri_membership-inference_2017}
\bibfield{author}{\bibinfo{person}{Reza Shokri}, \bibinfo{person}{Marco
  Stronati}, \bibinfo{person}{Congzheng Song}, {and} \bibinfo{person}{Vitaly
  Shmatikov}.} \bibinfo{year}{2017}\natexlab{}.
\newblock \showarticletitle{Membership Inference Attacks Against Machine
  Learning Models}. In \bibinfo{booktitle}{\emph{Security and Privacy}}.
  \bibinfo{publisher}{IEEE}.
\newblock


\bibitem[\protect\citeauthoryear{So, Güler, and Avestimehr}{So
  et~al\mbox{.}}{2020}]%
        {so_byzantine-resilient_2020}
\bibfield{author}{\bibinfo{person}{Jinhyun So}, \bibinfo{person}{Başak
  Güler}, {and} \bibinfo{person}{A.~Salman Avestimehr}.}
  \bibinfo{year}{2020}\natexlab{}.
\newblock \showarticletitle{Byzantine-Resilient Secure Federated Learning}.
\newblock \bibinfo{journal}{\emph{IEEE Journal on Selected Areas in
  Communications}}.
\newblock


\bibitem[\protect\citeauthoryear{So, Güler, and Avestimehr}{So
  et~al\mbox{.}}{2021}]%
        {so_turbo-aggregate_2021}
\bibfield{author}{\bibinfo{person}{Jinhyun So}, \bibinfo{person}{Başak
  Güler}, {and} \bibinfo{person}{A.~Salman Avestimehr}.}
  \bibinfo{year}{2021}\natexlab{}.
\newblock \showarticletitle{Turbo-Aggregate: Breaking the Quadratic Aggregation
  Barrier in Secure Federated Learning}.
\newblock \bibinfo{journal}{\emph{IEEE Journal on Selected Areas in Information
  Theory}}.
\newblock


\bibitem[\protect\citeauthoryear{Spooren, Preuveneers, Desmet, Janssen, and
  Joosen}{Spooren et~al\mbox{.}}{2019}]%
        {spooren_detection_2019}
\bibfield{author}{\bibinfo{person}{Jan Spooren}, \bibinfo{person}{Davy
  Preuveneers}, \bibinfo{person}{Lieven Desmet}, \bibinfo{person}{Peter
  Janssen}, {and} \bibinfo{person}{Wouter Joosen}.}
  \bibinfo{year}{2019}\natexlab{}.
\newblock \showarticletitle{Detection of Algorithmically Generated Domain Names
  Used by Botnets: A Dual Arms Race}. In \bibinfo{booktitle}{\emph{Symposium On
  Applied Computing}}. \bibinfo{publisher}{ACM}.
\newblock


\bibitem[\protect\citeauthoryear{Sun, Kairouz, Suresh, and McMahan}{Sun
  et~al\mbox{.}}{2019}]%
        {sun_can_2019}
\bibfield{author}{\bibinfo{person}{Ziteng Sun}, \bibinfo{person}{Peter
  Kairouz}, \bibinfo{person}{Ananda~Theertha Suresh}, {and}
  \bibinfo{person}{H~Brendan McMahan}.} \bibinfo{year}{2019}\natexlab{}.
\newblock \showarticletitle{Can You Really Backdoor Federated Learning?}
\newblock \bibinfo{journal}{\emph{arXiv:1911.07963}}.
\newblock


\bibitem[\protect\citeauthoryear{Tolpegin, Truex, Gursoy, and Liu}{Tolpegin
  et~al\mbox{.}}{2020}]%
        {tolpegin_data_2020}
\bibfield{author}{\bibinfo{person}{Vale Tolpegin}, \bibinfo{person}{Stacey
  Truex}, \bibinfo{person}{Mehmet~Emre Gursoy}, {and} \bibinfo{person}{Ling
  Liu}.} \bibinfo{year}{2020}\natexlab{}.
\newblock \showarticletitle{Data {Poisoning} {Attacks} {Against} {Federated}
  {Learning} {Systems}}. In \bibinfo{booktitle}{\emph{Computer Security --
  ESORICS 2020}}. \bibinfo{publisher}{Springer}.
\newblock


\bibitem[\protect\citeauthoryear{Wang, Song, Zhang, Song, Wang, and Qi}{Wang
  et~al\mbox{.}}{2019}]%
        {wang_beyond_2018}
\bibfield{author}{\bibinfo{person}{Zhibo Wang}, \bibinfo{person}{Mengkai Song},
  \bibinfo{person}{Zhifei Zhang}, \bibinfo{person}{Yang Song},
  \bibinfo{person}{Qian Wang}, {and} \bibinfo{person}{Hairong Qi}.}
  \bibinfo{year}{2019}\natexlab{}.
\newblock \showarticletitle{Beyond Inferring Class Representatives: User-Level
  Privacy Leakage From Federated Learning}. In
  \bibinfo{booktitle}{\emph{Computer Communications}}.
  \bibinfo{publisher}{IEEE}.
\newblock


\bibitem[\protect\citeauthoryear{Weiss, Khoshgoftaar, and Wang}{Weiss
  et~al\mbox{.}}{2016}]%
        {weiss_survey_2016}
\bibfield{author}{\bibinfo{person}{Karl Weiss}, \bibinfo{person}{Taghi~M.
  Khoshgoftaar}, {and} \bibinfo{person}{DingDing Wang}.}
  \bibinfo{year}{2016}\natexlab{}.
\newblock \showarticletitle{A survey of transfer learning}.
\newblock \bibinfo{journal}{\emph{Journal of Big Data}} \bibinfo{volume}{3},
  \bibinfo{number}{1}.
\newblock


\bibitem[\protect\citeauthoryear{Woodbridge, Anderson, Ahuja, and
  Grant}{Woodbridge et~al\mbox{.}}{2016}]%
        {woodbridge_predicting_2016}
\bibfield{author}{\bibinfo{person}{Jonathan Woodbridge},
  \bibinfo{person}{Hyrum~S. Anderson}, \bibinfo{person}{Anjum Ahuja}, {and}
  \bibinfo{person}{Daniel Grant}.} \bibinfo{year}{2016}\natexlab{}.
\newblock \showarticletitle{Predicting Domain Generation Algorithms with Long
  Short-Term Memory Networks}.
\newblock \bibinfo{journal}{\emph{arXiv:1611.00791}}.
\newblock


\bibitem[\protect\citeauthoryear{Yadav and Gupta}{Yadav and Gupta}{2021}]%
        {yadav_clustering_2021}
\bibfield{author}{\bibinfo{person}{Krishna Yadav} {and} \bibinfo{person}{B.~B.
  Gupta}.} \bibinfo{year}{2021}\natexlab{}.
\newblock \showarticletitle{Clustering {Algorithm} to {Detect} {Adversaries} in
  {Federated} {Learning}}.
\newblock \bibinfo{journal}{\emph{arXiv:2102.10799}}.
\newblock


\bibitem[\protect\citeauthoryear{Yadav and Reddy}{Yadav and Reddy}{2011}]%
        {yadav_winning_2012}
\bibfield{author}{\bibinfo{person}{Sandeep Yadav} {and}
  \bibinfo{person}{A.~L.~Narasimha Reddy}.} \bibinfo{year}{2011}\natexlab{}.
\newblock \showarticletitle{Winning with DNS Failures: Strategies for Faster
  Botnet Detection}. In \bibinfo{booktitle}{\emph{Security and Privacy in
  Communication Systems}}. Springer.
\newblock


\bibitem[\protect\citeauthoryear{Yang, Andrew, Eichner, Sun, Li, Kong, Ramage,
  and Beaufays}{Yang et~al\mbox{.}}{2018}]%
        {yang_applied_2018}
\bibfield{author}{\bibinfo{person}{Timothy Yang}, \bibinfo{person}{Galen
  Andrew}, \bibinfo{person}{Hubert Eichner}, \bibinfo{person}{Haicheng Sun},
  \bibinfo{person}{Wei Li}, \bibinfo{person}{Nicholas Kong},
  \bibinfo{person}{Daniel Ramage}, {and} \bibinfo{person}{Fran{\c{c}}oise
  Beaufays}.} \bibinfo{year}{2018}\natexlab{}.
\newblock \showarticletitle{Applied federated learning: Improving google
  keyboard query suggestions}.
\newblock \bibinfo{journal}{\emph{arXiv:1812.02903}}.
\newblock


\bibitem[\protect\citeauthoryear{Ye and Cui}{Ye and Cui}{2021}]%
        {ye_sample-based_2021}
\bibfield{author}{\bibinfo{person}{Chencheng Ye} {and} \bibinfo{person}{Ying
  Cui}.} \bibinfo{year}{2021}\natexlab{}.
\newblock \showarticletitle{Sample-based {Federated} {Learning} via
  {Mini}-batch {SSCA}}.
\newblock \bibinfo{journal}{\emph{arXiv:2103.09506}}.
\newblock


\bibitem[\protect\citeauthoryear{Yu, Pan, Hu, Nascimento, and De~Cock}{Yu
  et~al\mbox{.}}{2018}]%
        {yu_character_2018}
\bibfield{author}{\bibinfo{person}{Bin Yu}, \bibinfo{person}{Jie Pan},
  \bibinfo{person}{Jiaming Hu}, \bibinfo{person}{Anderson Nascimento}, {and}
  \bibinfo{person}{Martine De~Cock}.} \bibinfo{year}{2018}\natexlab{}.
\newblock \showarticletitle{Character Level Based Detection of {DGA} Domain
  Names}. In \bibinfo{booktitle}{\emph{{{International Joint Conference}} on
  {{Neural Networks}}}}. \bibinfo{publisher}{{IEEE}}.
\newblock


\bibitem[\protect\citeauthoryear{Zheng, Chen, Long, and Su}{Zheng
  et~al\mbox{.}}{2021}]%
        {zheng_federated_2021}
\bibfield{author}{\bibinfo{person}{Qinqing Zheng}, \bibinfo{person}{Shuxiao
  Chen}, \bibinfo{person}{Qi Long}, {and} \bibinfo{person}{Weijie Su}.}
  \bibinfo{year}{2021}\natexlab{}.
\newblock \showarticletitle{Federated f-Differential Privacy}. In
  \bibinfo{booktitle}{\emph{Artificial Intelligence and Statistics}}. PMLR.
\newblock


\bibitem[\protect\citeauthoryear{Zhu and Blaschko}{Zhu and Blaschko}{2021}]%
        {zhu_r-gap_2021}
\bibfield{author}{\bibinfo{person}{Junyi Zhu} {and} \bibinfo{person}{Matthew
  Blaschko}.} \bibinfo{year}{2021}\natexlab{}.
\newblock \showarticletitle{R-{GAP}: {Recursive} {Gradient} {Attack} on
  {Privacy}}.
\newblock \bibinfo{journal}{\emph{arXiv:2010.07733}}.
\newblock


\bibitem[\protect\citeauthoryear{Zhu, Liu, and Han}{Zhu et~al\mbox{.}}{2019}]%
        {zhu_deep_2019}
\bibfield{author}{\bibinfo{person}{Ligeng Zhu}, \bibinfo{person}{Zhijian Liu},
  {and} \bibinfo{person}{Song Han}.} \bibinfo{year}{2019}\natexlab{}.
\newblock \showarticletitle{Deep Leakage from Gradients}. In
  \bibinfo{booktitle}{\emph{Advances in Neural Information Processing
  Systems}}.
\newblock


\bibitem[\protect\citeauthoryear{Zhuang, Qi, Duan, Xi, Zhu, Zhu, Xiong, and
  He}{Zhuang et~al\mbox{.}}{2021}]%
        {zhuang_comprehensive_2021}
\bibfield{author}{\bibinfo{person}{Fuzhen Zhuang}, \bibinfo{person}{Zhiyuan
  Qi}, \bibinfo{person}{Keyu Duan}, \bibinfo{person}{Dongbo Xi},
  \bibinfo{person}{Yongchun Zhu}, \bibinfo{person}{Hengshu Zhu},
  \bibinfo{person}{Hui Xiong}, {and} \bibinfo{person}{Qing He}.}
  \bibinfo{year}{2021}\natexlab{}.
\newblock \showarticletitle{A Comprehensive Survey on Transfer Learning}.
\newblock \bibinfo{journal}{\emph{Proc. IEEE}} \bibinfo{volume}{109},
  \bibinfo{number}{1}.
\newblock


\end{thebibliography}
